\documentclass[12pt]{article}
\usepackage{amsmath,amssymb,axodraw}
\usepackage[dvips]{epsfig}
\usepackage[small]{caption}

\def\refeq#1{\mbox{(\ref{#1})}}
\def\reffi#1{\mbox{Fig.~\ref{#1}}}
\def\refta#1{\mbox{Table~\ref{#1}}}
\def\refse#1{\mbox{Section~\ref{#1}}}
\def\citere#1{\mbox{Ref.~\cite{#1}}}
\def\citeres#1{\mbox{Refs.~\cite{#1}}}

\def\limfunc#1{\mathop{\rm #1}}
\def\Re{\limfunc{Re}}
\def\Im{\limfunc{Im}}
\def\Li2{\limfunc{Li_2}}
\def\M{{\cal M}}
\def\O{{\cal O}}
\def\d{\mathrm d}
\def\i{\mathrm i}

\def\eps{\varepsilon}
\def\la{\lambda}
\def\de{\delta}
\def\cabs#1{\left| #1\right|}
\def\ie{i.e.\ }
\def\eg{e.g.\ }
\newcommand{\cs}{cross-section}

\newcommand{\Oa}{\mathswitch{{\cal{O}}(\alpha)}}
\newcommand{\Oaa}{\mathswitch{{\cal{O}}(\alpha^2)}}

\def\mathswitch#1{\relax\ifmmode#1\else$#1$\fi}
\def\mathswitchr#1{\relax\ifmmode{\mathrm{#1}}\else$\mathrm{#1}$\fi}
\newcommand{\PW}{\mathswitchr W}
\newcommand{\PZ}{\mathswitchr Z}

\newcommand{\PH}{\mathswitchr H}
\newcommand{\Pe}{\mathswitchr e}

\newcommand{\Pd}{\mathswitchr d}
\newcommand{\Pu}{\mathswitchr u}
\newcommand{\Ps}{\mathswitchr s}
\newcommand{\Pc}{\mathswitchr c}
\newcommand{\Pb}{\mathswitchr b}
\newcommand{\Pt}{\mathswitchr t}
\newcommand{\Pep}{\mathswitchr {e^+}}   
\newcommand{\Pem}{\mathswitchr {e^-}}
\newcommand{\PWp}{\mathswitchr {W^+}}
\newcommand{\PWm}{\mathswitchr {W^-}}

\newcommand{\MW}{\mathswitch {M_\PW}}

\newcommand{\MZ}{\mathswitch {M_\PZ}}
\newcommand{\MH}{\mathswitch {M_\PH}}
\newcommand{\Me}{\mathswitch {m_\Pe}}
\newcommand{\Mmy}{\mathswitch {m_\mu}}
\newcommand{\Mta}{\mathswitch {m_\tau}} 
\newcommand{\Md}{\mathswitch {m_\Pd}}
\newcommand{\Mu}{\mathswitch {m_\Pu}}
\newcommand{\Ms}{\mathswitch {m_\Ps}}
\newcommand{\Mc}{\mathswitch {m_\Pc}}
\newcommand{\Mb}{\mathswitch {m_\Pb}}
\newcommand{\Mt}{\mathswitch {m_\Pt}}
\newcommand{\sw}{\mathswitch {s_\PW}}
\newcommand{\cw}{\mathswitch {c_\PW}}
\newcommand{\WWWW}{\PWp\PWm\to\PWp\PWm}
\newcommand{\WLWLWLWL}{\PW_\rL^+\PW_\rL^-\to\PW_\rL^+\PW_\rL^-}
\newcommand{\pppp}{\varphi^+\varphi^-\to\varphi^+\varphi^-}

\newcommand{\ee}{\Pep\Pem}
\newcommand{\ZZZZ}{\PZ\PZ\to\PZ\PZ}
\newcommand{\born}{\mathrm{Born}}
\newcommand{\soft}{\mathrm{soft}}

\newcommand{\rL}{{\mathrm{L}}}
\newcommand{\rT}{{\mathrm{T}}}
\newcommand{\TeV}{\unskip\,\mathrm{TeV}}
\newcommand{\GeV}{\unskip\,\mathrm{GeV}}
\newcommand{\MeV}{\unskip\,\mathrm{MeV}}
\newcommand{\FA}{{\sl Feyn\-Arts}}
\newcommand{\mma}{{\sl Mathematica}}
\newcommand{\FO}{{\sl Form}}
\def\thcut{\theta_{\text{cut}}}

\newcommand{\cpc}[3]{{\sl Comp. Phys. Commun.} {\bf #1} (19#2) #3}
\newcommand{\fp}[3]{{\sl Fortschr. Phys.} {\bf #1} (19#2) #3}
\newcommand{\nc}[3]{{\sl Nuovo Cimento} {\bf #1} (19#2) #3}
\newcommand{\np}[3]{{\sl Nucl. Phys.} {\bf #1} (19#2)~#3}
\newcommand{\pl}[3]{{\sl Phys. Lett.} {\bf #1} (19#2) #3}
\newcommand{\pr}[3]{{\sl Phys. Rev.} {\bf #1} (19#2) #3}
\newcommand{\prl}[3]{{\sl Phys. Rev. Lett.} {\bf #1} (19#2) #3}
\newcommand{\sptp}[3]{{\sl Suppl. Prog. Theor. Phys.} {\bf #1} (19#2) #3}
\newcommand{\zp}[3]{{\sl Z. Phys.} {\bf #1} (19#2) #3}
\newcommand{\vj}[4]{{\sl #1~}{\bf #2} (19#3) #4}

\makeatletter

\def\section{\@startsection {section}{1}{\z@}{+3.0ex plus +1ex minus
  +.2ex}{2.3ex plus .2ex}{\normalsize\bf\boldmath}}
\def\subsection{\@startsection{subsection}{2}{\z@}{+2.5ex plus +1ex
minus +.2ex}{1.5ex plus .2ex}{\normalsize\bf\boldmath}}
\def\subsubsection{\@startsection{subsubsection}{3}{\z@}{+3.25ex plus
 +1ex minus +.2ex}{1.5ex plus .2ex}{\normalsize\bf\boldmath}}

\newcount\@tempcntc
\def\@citex[#1]#2{\if@filesw\immediate\write\@auxout{\string\citation{#2}}\fi
  \@tempcnta\z@\@tempcntb\m@ne\def\@citea{}\@cite{\@for\@citeb:=#2\do
    {\@ifundefined
       {b@\@citeb}{\@citeo\@tempcntb\m@ne\@citea
        \def\@citea{,\penalty\@m\ }{\bf ?}\@warning
       {Citation `\@citeb' on page \thepage \space undefined}}%
    {\setbox\z@\hbox{\global\@tempcntc0\csname
b@\@citeb\endcsname\relax}%
     \ifnum\@tempcntc=\z@ \@citeo\@tempcntb\m@ne
       \@citea\def\@citea{,\penalty\@m}
       \hbox{\csname b@\@citeb\endcsname}%
     \else
      \advance\@tempcntb\@ne
      \ifnum\@tempcntb=\@tempcntc
      \else\advance\@tempcntb\m@ne\@citeo
      \@tempcnta\@tempcntc\@tempcntb\@tempcntc\fi\fi}}\@citeo}{#1}}

\def\@citeo{\ifnum\@tempcnta>\@tempcntb\else\@citea
  \def\@citea{,\penalty\@m}%
  \ifnum\@tempcnta=\@tempcntb\the\@tempcnta\else
   {\advance\@tempcnta\@ne\ifnum\@tempcnta=\@tempcntb \else
\def\@citea{--}\fi
    \advance\@tempcnta\m@ne\the\@tempcnta\@citea\the\@tempcntb}\fi\fi}

\makeatother

\begin{document}

\thispagestyle{empty}
\def\thefootnote{\fnsymbol{footnote}}
\setcounter{footnote}{1}
\null
\strut\hfill PSI-PR-97-31\\
\strut\hfill KA-TP-16-1997\\
\strut\hfill hep-ph/9711302
\vskip 0cm
\vfill
\begin{center}
{\Large\bf 
\boldmath{Radiative Corrections to $\WWWW$ \\[1ex]
in the Electroweak Standard Model}}
\vskip 2.5em
{\large\sc A.~Denner}\\[1ex]
{\normalsize\it Paul-Scherrer-Institut, W\"urenlingen und Villigen\\
CH-5232 Villigen PSI, Switzerland}\\[2ex]
{\large\sc T.~Hahn} \\[1ex]
{\normalsize\it Institut f\"ur Theoretische Physik, Universit\"at
Karlsruhe\\
D-76128 Karlsruhe, Germany}
\par \vskip 1em
\end{center} \par
\vskip 1cm 
\vfill
{\bf Abstract:} \par
The cross-section for $\WWWW$ with arbitrarily polarized W bosons is
calculated within the Electroweak Standard Model including the complete
virtual and soft-photonic $\Oa$ corrections. We show the numerical
importance of the radiative corrections for the dominating polarized
cross-sections and for the unpolarized cross-section. The numerical
accuracy of the equivalence theorem is investigated in $\Oa$ by comparing
the cross-section for purely longitudinal W~bosons obtained from the
equivalence theorem and from the direct calculation. We point out that
the instability of the \PW~boson, which is inherent in the one-loop
corrections, prevents a consistent calculation of radiative
corrections to the scattering of on-real-mass-shell longitudinal
W~bosons beyond $\Oa$. 
\par
\vskip 1cm 
\noindent
November 1997\par
\null
\setcounter{page}{0}
\clearpage
\def\thefootnote{\arabic{footnote}}
\setcounter{footnote}{0}


\section{Introduction}

Gauge-boson scattering provides a direct way to probe both the
non-abelian and scalar sector of the Electroweak Standard Model (SM).
Since in lowest order all gauge-boson scattering amplitudes involve
only interactions between gauge and scalar bosons, the corresponding
cross-sections depend very sensitively on the non-abelian and scalar
sector of the underlying theory. This sensitivity is even enhanced for
high-energetic, longitudinally polarized, massive gauge bosons owing
to the presence of gauge cancellations. For these reasons, gauge-boson
scattering has found continuous interest in the literature
\cite{DiM73,Pa85,Da89,Bo90} since the early years of spontaneously
broken gauge theories. The sensitivity of gauge-boson scattering to
the non-abelian gauge couplings has been investigated, for instance, in
\citere{Ba93}.

Gauge-boson scattering can be studied in principle at all high-energy
colliders, such as pp colliders like the LHC \cite{Ba94}, $\ee$ colliders
like the proposed NLC \cite{Ze92}, or $\mu^+\mu^-$ colliders
\cite{BaBGH97}. These reactions naturally appear as subprocesses, e.g.\ as
final-state interactions in hadron or lepton collisions. At high energies
($E\gg\MW$) the incoming particles radiate plenty of gauge bosons. Similar
to the well-known Weizs\"acker--Williams approximation for photonic
reactions also massive vector-boson scattering at high energies can be
approximated by convoluting the vector-boson cross-section with the
corresponding flux of vector bosons. This approximation is known as the
{\em equivalent vector-boson method} (see e.g.\ \citere{KuS96} and
references therein).

Gauge-boson scattering reactions have been studied at tree level in
\citere{DiM73}. So far, only the leading radiative corrections in the
limit of high energies and large Higgs-boson masses ($E,\MH\gg\MW$) have
been calculated \cite{Pa85,Da89,Bo90}. In most of these calculations the
scattering of only longitudinal gauge bosons was considered, and the
Goldstone-boson equivalence theorem \cite{CoLT74} was used. A complete
$\Oa$ calculation was available only for the process
$\gamma\gamma\to\PWp\PWm$ \cite{DeDS95,Ji97}.

In a recent paper \cite{DeDH97} the complete one-loop electroweak
radiative corrections to the simplest massive gauge-boson scattering
process, $\ZZZZ$, have been discussed. By means of this example
many characteristic features of massive gauge-boson scattering processes 
could be studied at the one-loop level, such as the treatment of the
Higgs-boson resonance and the gauge cancellations, which are 
quantitatively expressed by the Goldstone-boson equivalence theorem. 

In this paper we perform a similar study of elastic $\PWp\PWm$ scattering,
which is one of the dominating gauge-boson scattering processes at all
energies. Because the \PW~bosons are charged, real bremsstrahlung
corrections are required that lead to additional complications. Since we
are mainly interested in the weak corrections, we include these
corrections only in the soft-photon limit and omit the radiation of hard
photons that depends on the experimental situation. Moreover, we restrict
our discussion to the case of a relatively light Higgs boson so that no
Higgs resonance occurs. Such a light Higgs boson is favoured by precision
experiments \cite{LEPEWWG96}. Furthermore, the treatment of the Higgs
resonance requires to include higher-order corrections. But as we show in
this paper, the calculation of corrections of order $\alpha^2$ or higher
for the scattering of on-real-mass-shell longitudinal \PW~bosons becomes
inconsistent owing to the finite width of the \PW~bosons, which enters
unavoidably via loop diagrams.

This paper is organized as follows: After some preliminary remarks 
about kinematics, conventions, and discrete symmetries in \refse{se:prelim},
we discuss the lowest-order cross-sections in \refse{se:locs}. In
\refse{se:rcs} we describe the explicit calculation and the structure of
the $\Oa$ corrections. Numerical results are presented in
\refse{se:numres}. Section \ref{sect:problems} is devoted to a discussion
of problems that occur in the calculation of the radiative corrections to
$\WWWW$ beyond $\Oa$. Section~\ref{se:concl} contains our conclusions. 


\section{Kinematics, notation, and discrete symmetries}
\label{se:prelim}

We consider the reaction 
\begin{equation}
\PW^+(k_1,\lambda_1)+\PW^-(k_2,\lambda_2)\to
\PW^+(k_3,\lambda_3)+\PW^-(k_4,\lambda_4)\,,
\end{equation}
where $k_i$ and $\lambda_i$ represent the momenta and helicities of the
W bosons, respectively. In the following, the indices L, T, and U denote
longitudinal ($\lambda_i=0$), transverse ($\lambda_i=\pm$) and unpolarized
W bosons, respectively, and definite polarization combinations are
labelled by a sequence of four letters, e.g.\ LTLT stands for
$\mathswitchr{W^+_L}\mathswitchr{W^-_T}\to\mathswitchr{W^+_L}
\mathswitchr{W^-_T}$.

The incoming particles travel along the $z$ axis and are scattered into
the $x$--$z$ plane. In the centre-of-mass system (CMS) the momenta and
polarization vectors $\eps_i(\lambda_i)$ explicitly read
\begin{equation}
\label{eq:polvecs}
\begin{aligned}
k_1^\mu &= \left(E,\,0,\,0,\,-p\right), &
k_3^\mu &= \left(E,\,-p\sin \theta ,\,0,\,-p\cos\theta\right), \\
\eps_1^\mu(0) &= \left(-p,\,0,\,0,\,E \right)/\MW, & \qquad
\eps_3^{\mu,\ast}(0) &=
  \left(p,\,-E\sin\theta,\,0,\,-E\cos\theta\right)/\MW, \\
\eps_1^\mu(\pm) &= \left(0,\,-1,\,\pm\i,\,0\right)/\sqrt{2}\,, &
\eps_3^{\mu,\ast}(\pm) &=
  \left(0,\,-\cos\theta,\,\mp\i,\,\sin\theta\right)/\sqrt{2}\,,\\[1ex]
k_2^\mu &= \left(E,\,0,\,0,\,p\right), &
k_4^\mu &= \left(E,\,p\sin\theta,\,0,\,p\cos\theta\right), \\
\eps_2^\mu(0) &= \left(-p,\,0,\,0,\,-E\right)/\MW, &
\eps_4^{\mu,\ast}(0) &=
  \left(p,\,E\sin\theta,\,0,\,E\cos\theta\right)/\MW, \\
\eps_2^\mu(\pm) &= \left(0,\,1,\,\pm\i,\,0\right)/\sqrt{2}\,, &
\eps_4^{\mu,\ast}(\pm) &=
  \left(0,\,\cos\theta,\,\mp\i,\,-\sin\theta\right)/\sqrt{2}
\end{aligned}
\end{equation}
in terms of the energy $E$ of the W bosons, their momentum 
$p=\sqrt{E^2-\MW^2}$, and the scattering angle $\theta$. 
The Mandelstam variables are defined as
\begin{equation}
\begin{aligned}
s &= (k_1+k_2)^2=4E^2\,, \\
t &= (k_1-k_3)^2=-4p^2\sin^2\theta/2\,, \\
u &= (k_1-k_4)^2=-4p^2\cos^2\theta/2\,.
\end{aligned}
\end{equation}

The polarized differential cross-section is obtained from the invariant
matrix element $\M$ as
\begin{equation}
\label{eq:diffWQ}
\left(\frac{\d\sigma}{\d\Omega}\right)_{\lambda_1\lambda_2\lambda_3
\lambda_4}=\frac{1}{64\pi^2s}\cabs{\M_{\lambda_1\lambda_2\lambda_3
\lambda_4}}^2,
\end{equation}
and the unpolarized cross-section results from averaging over the
polarizations of the incoming and summing over the polarizations of the 
outgoing particles,
\begin{equation}
\left(\frac{\d\sigma}{\d\Omega}\right)_{\text{unpol}}=
\frac 19\sum_{\lambda_1,\lambda_2=-1}^1\,
\sum_{\lambda_3,\lambda_4=-1}^1
\left(\frac{\d\sigma}{\d\Omega}\right)_{\lambda_1\lambda_2\lambda_3
\lambda_4}\,.
\end{equation}
More generally, the correct average is obtained by multiplying with 1/3
for each unpolarized W boson and by 1/2 for each transverse W boson in the
initial state.

The integrated cross-section is defined by
\begin{equation}
\sigma=\int_{0^\circ}^{360^\circ}\d\varphi
\int_{\thcut}^{180^\circ-\thcut}
\d\theta\,\sin\theta\,\frac{\d\sigma}{\d\Omega}\,,
\end{equation}
where $\thcut$ denotes an angular cut which is set to $10^\circ$ in the
numerical calculations.

Discrete symmetries relate different helicity amplitudes. CPT symmetry
implies 
\begin{equation}
\label{CPTrel}
{\cal M}_{\lambda_1 \lambda_2 \lambda_3 \lambda_4} =
{\cal M}_{\lambda_3 \lambda_4 \lambda_1 \lambda_2}\,.
\end{equation}
Since we 
neglect quark mixing,
also CP is
conserved
and as a consequence, the helicity amplitudes are related as
follows
\begin{equation}
{\cal M}_{\lambda_1 \lambda_2 \lambda_3 \lambda_4} =
{\cal M}_{-\lambda_2 -\lambda_1 -\lambda_4 -\lambda_3}\,.
\end{equation}
These symmetries reduce the number of independent polarized amplitudes
from 81 to 27.


\section{Lowest-order cross-sections}
\label{se:locs}

\subsection{General properties}
\label{se:gp}

\begin{figure}
\begin{center}
\epsfig{figure=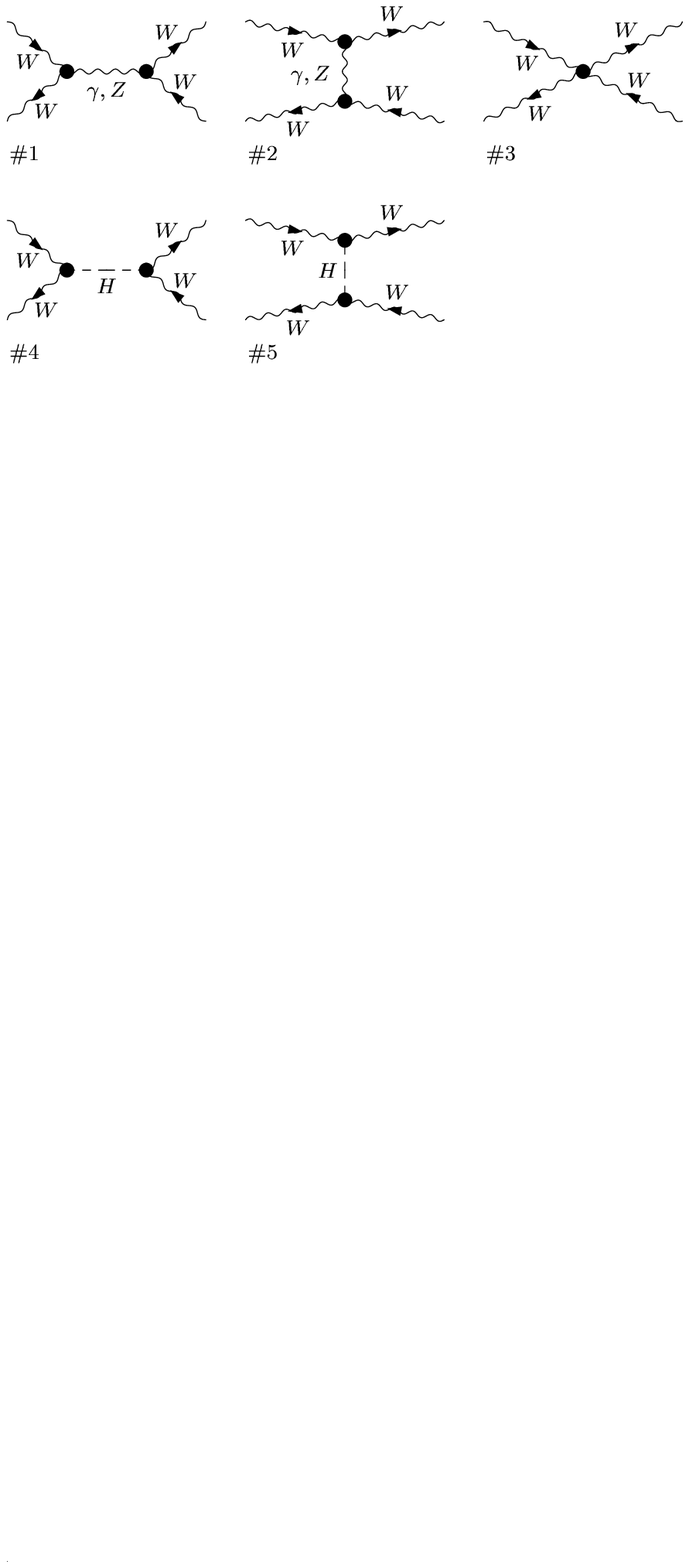,width=.6\linewidth}
\end{center}
\vspace*{-2ex}
\caption{\label{fig:borndiags}The lowest-order diagrams contributing to
$\WWWW$.}
\end{figure}
In lowest order, seven Feynman diagrams exist for $\WWWW$ as shown
in \reffi{fig:borndiags}. Owing to charge conservation there are no
$u$-channel diagrams, and therefore the cross-section is expected to have
a forward--backward asymmetry. The resulting tree-level amplitude reads
\begin{equation}
\label{eq:born}
\begin{aligned}
\M_\born=\,
& -e^2\left(\frac{1}{s}+\frac{\cw^2}{\sw^2}\frac{1}{s-\MZ^2}\right)\M_1
-e^2\left(\frac{1}{t}+\frac{\cw^2}{\sw^2}\frac{1}{t-\MZ^2}\right)\M_2\\
& {}-\frac{e^2}{\sw^2}\M_3
-\frac{e^2\MW^2}{\sw^2}\frac{1}{s-\MH^2}\,\M_4 
-\frac{e^2\MW^2}{\sw^2}\frac{1}{t-\MH^2}\,\M_5\,.
\end{aligned}
\end{equation}
The matrix elements $\M_1$--$\M_5$ correspond to the diagrams
\#1--\,\#5 in \reffi{fig:borndiags}, respectively, and contain the
complete kinematical information and polarization dependence,
\begin{align}
\M_1=\:\:& (\eps_1\cdot\eps_2)(\eps_3^*\cdot\eps_4^*)(t-u) \notag \\
        &{}+4\Bigl\{(\eps_1\cdot\eps_2)\bigl[
          (\eps_3^*\cdot k_4)(\eps_4^*\cdot k_2)-
          (\eps_3^*\cdot k_2)(\eps_4^*\cdot k_3)\bigr] \notag \\
        &\; {}+(\eps_3^*\cdot\eps_4^*)\bigl[
          (\eps_1\cdot k_3)(\eps_2\cdot k_4)-
          (\eps_1\cdot k_4)(\eps_2\cdot k_3)\bigr]
          \vphantom{\Bigr\{} \\
        &\; {}-(\eps_2\cdot k_1)\bigl[
          (\eps_1\cdot\eps_3^*)(\eps_4^*\cdot k_3)-
          (\eps_1\cdot\eps_4^*)(\eps_3^*\cdot k_4)\bigr]
          \vphantom{\Bigr\{} \notag \\
        &\; {}-(\eps_1\cdot k_2)\bigl[
          (\eps_2\cdot\eps_4^*)(\eps_3^*\cdot k_4)-
          (\eps_2\cdot\eps_3^*)(\eps_4^*\cdot k_3)\bigr]\Bigr\}\,,
        \notag \displaybreak[0] \\
\M_2=\:\:& (\eps_1\cdot\eps_3^*)(\eps_2\cdot\eps_4^*)(s-u) \notag \\
        &{}-4\Big\{
          (\eps_1\cdot\eps_3^*)\bigl[
          (\eps_2\cdot k_3)(\eps_4^*\cdot k_2)+
          (\eps_2\cdot k_4)(\eps_4^*\cdot k_3)\bigr]
          \vphantom{\Bigr\{} \notag \\
        &\; {}+(\eps_2\cdot\eps_4^*)\bigl[
          (\eps_1\cdot k_2)(\eps_3^*\cdot k_4)-
          (\eps_1\cdot k_4)(\eps_3^*\cdot k_2)\bigr]          
          \vphantom{\Bigr\{} \\
        &\; {}-(\eps_1\cdot k_3)\bigl[
          (\eps_3^*\cdot\eps_4^*)(\eps_2\cdot k_4)+
          (\eps_2\cdot\eps_3^*)(\eps_4^*\cdot k_2)\bigr] 
          \vphantom{\Bigr\{}\notag \\
        &\; {}-(\eps_3^*\cdot k_1)\bigl[
          (\eps_1\cdot\eps_4^*)(\eps_2\cdot k_4)+
          (\eps_1\cdot\eps_2)(\eps_4^*\cdot k_2)\bigr]
          \Bigr\}\,,  \notag \displaybreak[0] \\
\M_3=\:\:& (\eps_1\cdot\eps_2)(\eps_3^*\cdot\eps_4^*)+
        (\eps_1\cdot\eps_3^*)(\eps_2\cdot\eps_4^*)-
        2(\eps_1\cdot\eps_4^*)(\eps_2\cdot\eps_3^*)\,,
        \vphantom{\Bigr\{}\displaybreak[0] \\
\M_4=\:\:& (\eps_1\cdot\eps_2)(\eps_3^*\cdot\eps_4^*) \,,
        \vphantom{\Bigr\{}\displaybreak[0] \\
\M_5=\:\:& (\eps_1\cdot\eps_3^*)(\eps_2\cdot\eps_4^*)\,.
\end{align}

For equal polarizations in the initial and final state, \ie for
$\la_1=\la_3$ and $\la_2=\la_4$, the lowest-order amplitude diverges for
small momentum transfer $t$ owing to the $t$-channel photon-exchange 
diagram. In fact, in this limit the lowest-order cross-section turns
into the Rutherford cross-section $\propto 1/t^2\propto
1/[p\sin(\theta/2)]^4$. If either $\la_1=\la_3$ or $\la_2=\la_4$, the
cross-section has at most a $1/t$ singularity for small $t$, and for
$\la_1\ne\la_3$ and $\la_2\ne\la_4$ it is regular.

\begin{figure}[p]
\begin{center}
\epsfig{figure=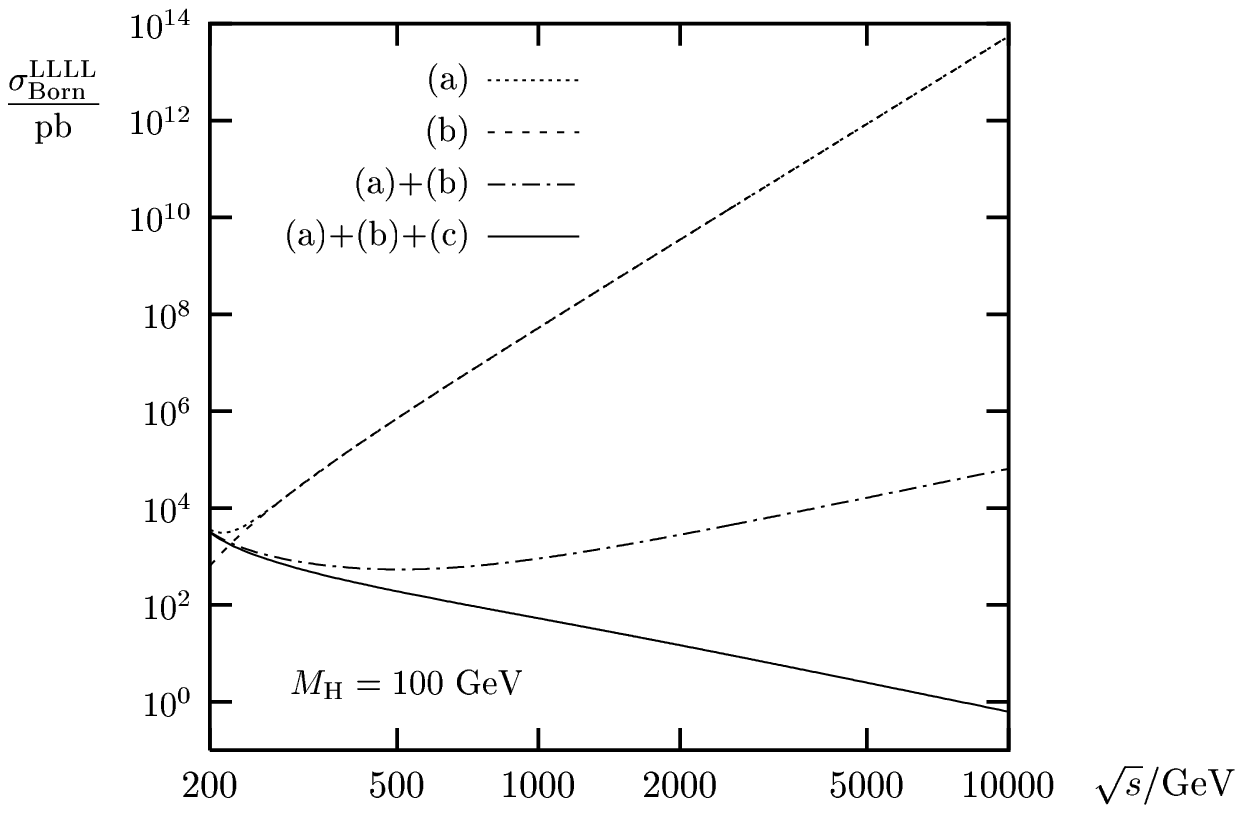}
\end{center}
\caption{\label{fig:unitarity}The cross-sections for longitudinal 
gauge-boson scattering resulting from subsets of the tree-level diagrams:
(a) diagrams involving only three-gauge-boson couplings,
(b) diagram involving only four-gauge-boson couplings,
(c) diagrams involving Higgs bosons.} 
\end{figure}

Each longitudinal \PW~boson enhances the individual matrix elements
$\M_1$--$\M_5$ by a factor $\sqrt s/\MW$. The cancellations guaranteed by
gauge symmetry prevent the cross-section for longitudinal $\PW\PW$
scattering from grossly violating unitarity at high energies. The gauge
cancellations are demonstrated graphically in \reffi{fig:unitarity} for
the total cross-section. The cross-section calculated from individual
diagrams involving three- (a) or four-gauge-boson couplings (b) grows with
$s^3$ for large $s$. Adding up all pure gauge-boson diagrams the leading
terms cancel resulting in a cross-section $\propto s$. Including the
Higgs-exchange diagrams, further cancellations take place such that the SM
cross-section drops as $1/s$. The cancellations are already quite
substantial at a few hundred GeV. 

The unitarity cancellations ensure that all polarized cross-sections 
decrease at least as $1/s$ at high energies. The cross-sections with an
odd number of longitudinal \PW~bosons are suppressed by an additional
factor $\MW^2/s$. The behaviour of various polarized cross-sections at
high energies is listed in \refta{tab:bornint}. 

\def\LOG{L_c}
\def\colstrut{\vrule width0pt height4ex depth3ex}
\begin{table}
\begin{small}
$$
\begin{array}{|l|c|c|} \hline
\multicolumn{1}{|c|}{\text{Polarization}} &
\sigma_{\text{Born}}=\pi\alpha^2/\sw^4\cdot(\text{entry}) \\
\hline\hline
\begin{array}{l}
\rL\rL\rL\rL
\end{array} &
        \dfrac 1s\left\{
        \dfrac{(\MH^2+\MZ^2)(2\MZ^2\LOG+\MH^2c)}{4\MW^4}
        +\dfrac{c\,(75-26c^2-c^4)}{48\cw^4(1-c^2)}\right\}
        \colstrut \\ \hline
 &
        -\dfrac 1{s^2}\biggl\{\dfrac{(\MH^2+4\MW^2+\MZ^2)^2\LOG}
        {4\MW^2}
        \hfill\colstrut \\
\raisebox{3.5ex}[0cm][0cm]{$
\begin{array}{l}
\rL\rL\rL\rT \\
\;=\rL\rL\rT\rL \\
\;=2\,\rL\rT\rL\rL \\
\;=2\,\rT\rL\rL\rL
\end{array}$} &
        \hfill{}
        +\dfrac{c\bigl[3\left(\MH^2+2(3\MW^2+\MZ^2)\right)^2
        -(6-c^2)(2\MW^2+\MZ^2)^2\bigr]}{12\MW^2}\biggr\}
        \colstrut \\ \hline
\begin{array}{l}
\rL\rL\rT\rT \\
\;=4\,\rT\rT\rL\rL
\end{array} &
        \dfrac 1s\dfrac{c\,(3+c^2)}{6}
        \colstrut \\ \hline
\begin{array}{l}
\rL\rT\rL\rT \\
\;=\rT\rL\rT\rL
\end{array} &
        \dfrac 1s\left\{\dfrac{c\,(5-c^2)}{1-c^2}+2\LOG\right\}
        \colstrut \\ \hline
 &
        \dfrac 1{s^3}\biggl\{\bigl[\MH^4+\MZ^4+
        2\MH^2(8\MW^2+\MZ^2)\bigr]\!\LOG+
        \dfrac{c\,(5-c^2)(\MH^4+\MZ^4)}{2(1-c^2)}
        \colstrut \\
\raisebox{3.5ex}[0cm][0cm]{$
\begin{array}{l}
\rL\rT\rT\rL \\
\;=\rT\rL\rL\rT
\end{array}$} &
        \hfill{}
        +\dfrac{c\,(8\MW^2+\MZ^2)\bigl[18\MH^2+(3+c^2)(8\MW^2+\MZ^2)
        \bigr]}{6}\biggr\}
        \colstrut \\ \hline
\begin{array}{l}
\rL\rT\rT\rT \\
\;=\rT\rL\rT\rT \\
\;=2\,\rT\rT\rL\rT \\
\;=2\,\rT\rT\rT\rL
\end{array} &
        -\dfrac 1{s^2}\MW^2(29c+3c^3+20\LOG)
        \colstrut \\ \hline
\begin{array}{l}
\rT\rT\rT\rT
\end{array} &
        \dfrac 1s\left\{\dfrac{c\,(75-26c^2-c^4)}{3(1-c^2)}
        +8\LOG\right\}
        \colstrut \\ \hline
\end{array}
$$
\end{small}
\caption{\label{tab:bornint}Leading terms for polarized integrated
lowest-order cross-sections in the high-energy limit. Here $c=\cos\thcut$,
$\LOG=\ln\frac{1-c}{1+c}$, and the relations between different
polarizations in the left column are to be read, for instance, as
$\sigma_{\rL\rT\rT\rT}=2\sigma_{\rT\rT\rL\rT}$, where the numerical
prefactors originate from the different spin averages.}
\end{table}

\subsection{Numerical results}

All following numerical results are obtained for an angular cut-off of
$\thcut=10^\circ$ and for a Higgs-boson mass $\MH=100\GeV$. This value is
within the range allowed by precision data, which favour a light Higgs
boson \cite{LEPEWWG96}. Since it lies below the threshold for $\WWWW$ no
Higgs resonance occurs. 

\begin{figure}
\begin{center}
\epsfig{figure=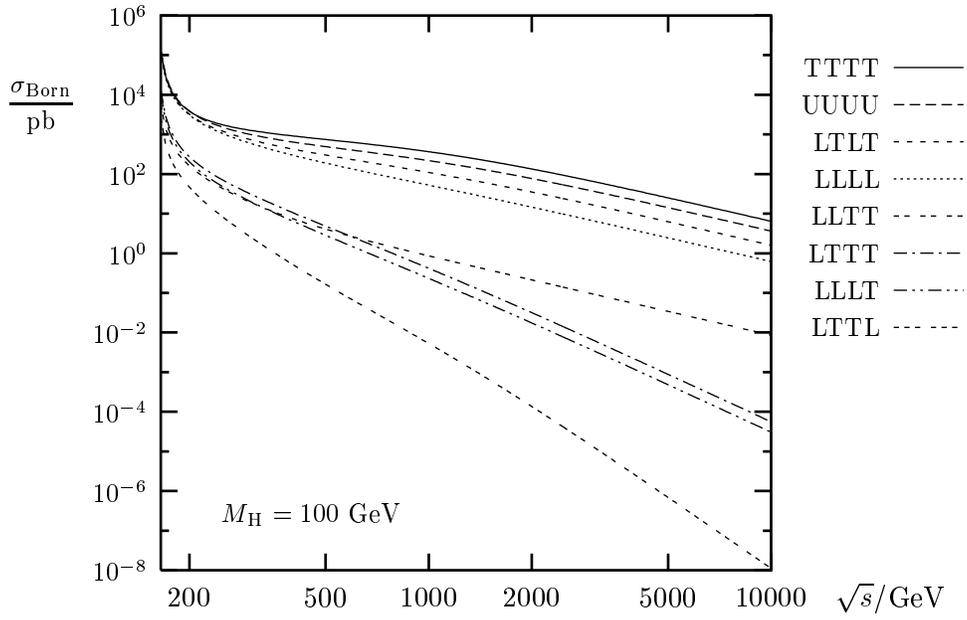}
\end{center}
\caption{\label{fig:borntotplots}The integrated lowest-order
cross-sections for various polarizations.}
\end{figure}
In \reffi{fig:borntotplots} the integrated lowest-order cross-section is
displayed for various \PW~polarizations. The curves nicely reflect the
high-energy behaviour of the analytical expressions in
\refta{tab:bornint}. The cross-sections for the polarizations LLLL, TTTT,
LTLT, and LTTL behave as $1/s$ at high energies, all others are suppressed
by additional factors $\MW^2/s$. Owing to the $1/t$ pole, the 
cross-sections with equal initial- and final-state polarizations, LLLL,
TTTT, and LTLT, are enhanced by a factor $1/(1-\cos^2\thcut)\approx 33$
for $\thcut=10^\circ$ and are therefore dominant. For these polarizations
the contribution of the backward hemisphere to the integrated 
cross-section is less than 1\% for energies above $1\TeV$ and at most
3\% for lower energies. Note that the size of the corresponding integrated
cross-sections strongly depends on the angular cut, $\thcut$, because of
the $t$-channel singularity. 

\begin{figure}
\begin{center}
\begin{tabular}{l}
\epsfig{figure=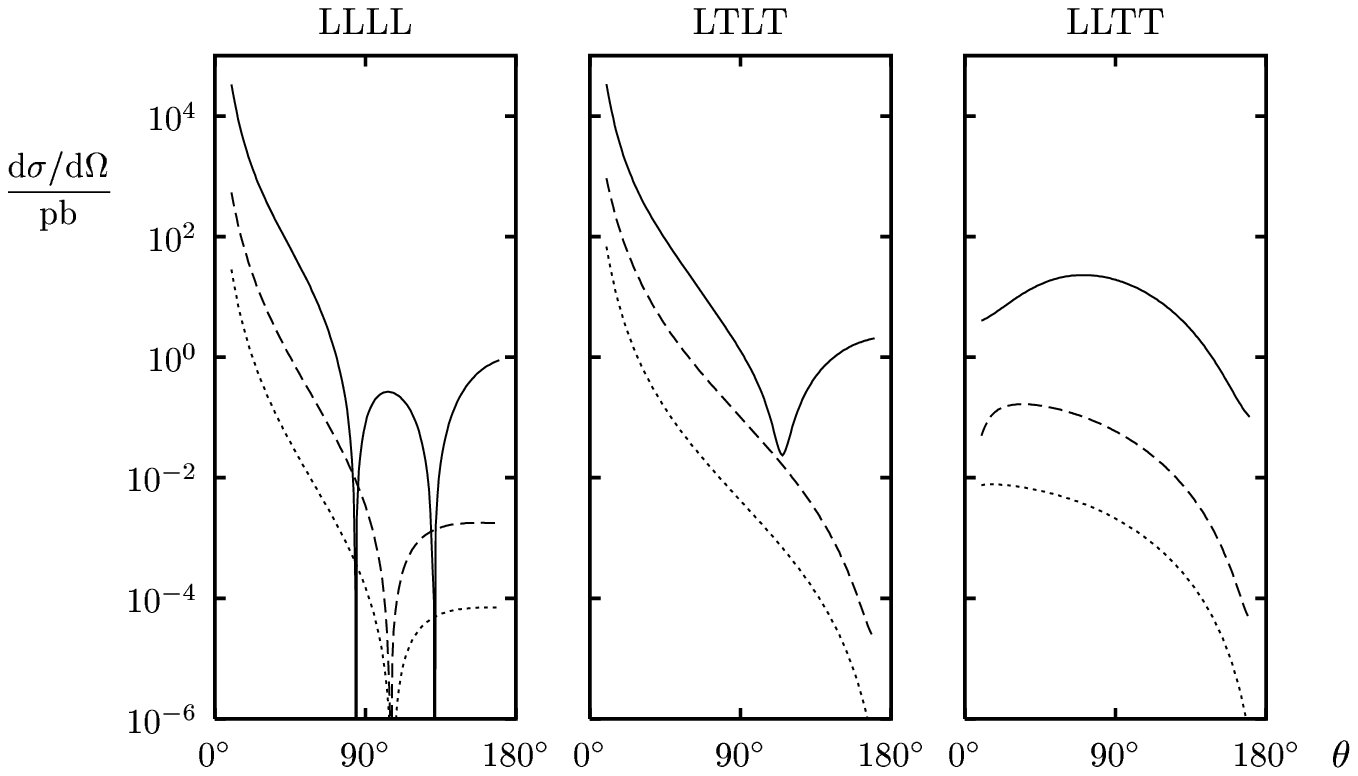} \\[1ex]
\epsfig{figure=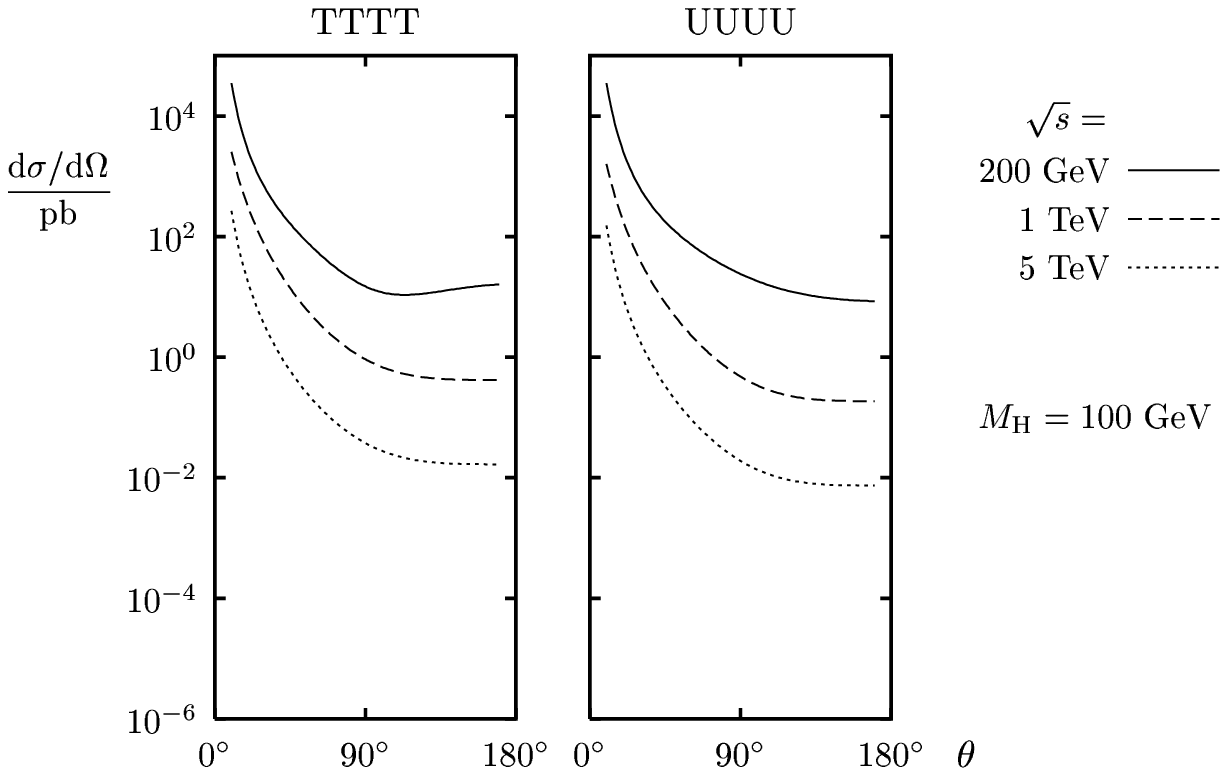}
\end{tabular}
\end{center}
\caption{\label{fig:borndiffplots}The differential lowest-order
cross-sections for various polarizations at different energies.}
\end{figure}
In \reffi{fig:borndiffplots} the differential lowest-order cross-sections
are shown for $\sqrt s=200\GeV$, $1\TeV$, and $5\TeV$ for the dominating
polarizations LLLL, LTLT, LLTT, TTTT, and for the unpolarized case UUUU.
For all these polarizations except LLTT the curves display the strong
forward peak resulting from the $t$-channel pole. The figures show that
the main contributions to the unpolarized cross-section in backward
direction come from completely transversally polarized bosons. The sharp
dips in the longitudinal cross-section correspond to zeros in the
amplitude. To be precise, the longitudinal cross-section is a rational
function in $t$ and has four such zeros. For $\sqrt{s}\lesssim 274\GeV$
two of these zeros and for $\sqrt{s}\gtrsim 274\GeV$ one of them lie in
the physical region. 


\section{$\Oa$ corrections}
\label{se:rcs}

\subsection{Calculational framework}
\label{se:calframe}

We have performed the calculation of the $\Oa$ radiative corrections in 
't~Hooft--Feynman gauge both in the conventional formalism and in the
background-field formalism. Ultraviolet (UV) divergences are treated
within dimensional regularization. We use the on-shell renormalization
scheme \cite{RoT73}, following the formulation worked out in \citere{De93}
for the conventional formalism and in \citere{DeDW95} for the
background-field formalism. In the conventional formalism the field
renormalization is fixed such that no external wave-function
renormalization is needed. In the renormalization scheme introduced in
\citere{DeDW95} for the background-field method the field renormalization
is determined by gauge invariance, and a non-trivial external
wave-function renormalization is required, as explicitly described in
\citere{DeD96}.

Our calculation is based on the second of the two methods described in
\citere{DeDH97}. The Feynman diagrams are generated with \FA\ 
\cite{KuBD91}. The resulting amplitudes are algebraically simplified with
a combination of \FO\ \cite{Ve91} and \mma\ and are then converted into a
{\sl Fortran} program. The tensor integrals are numerically reduced to
scalar integrals, which are evaluated using the {\sl FF} package
\cite{vOV90}. 

\subsection{Inventory}
 
At one-loop level, roughly 1000 diagrams contribute to $\WWWW$. The
diagrams can be classified into self-energy corrections, vertex
corrections, and box corrections. The self-energy corrections and the
vertex corrections can be further divided into $s$- and $t$-channel
contributions, the box corrections involve in addition $u$-channel
contributions. The $t$-channel contributions can be obtained from the
$s$-channel contributions via crossing, \ie via the interchange of the
outgoing $\PWm$~boson with the incoming $\PWp$~boson.

\begin{figure}
\begin{center}
\epsfig{figure=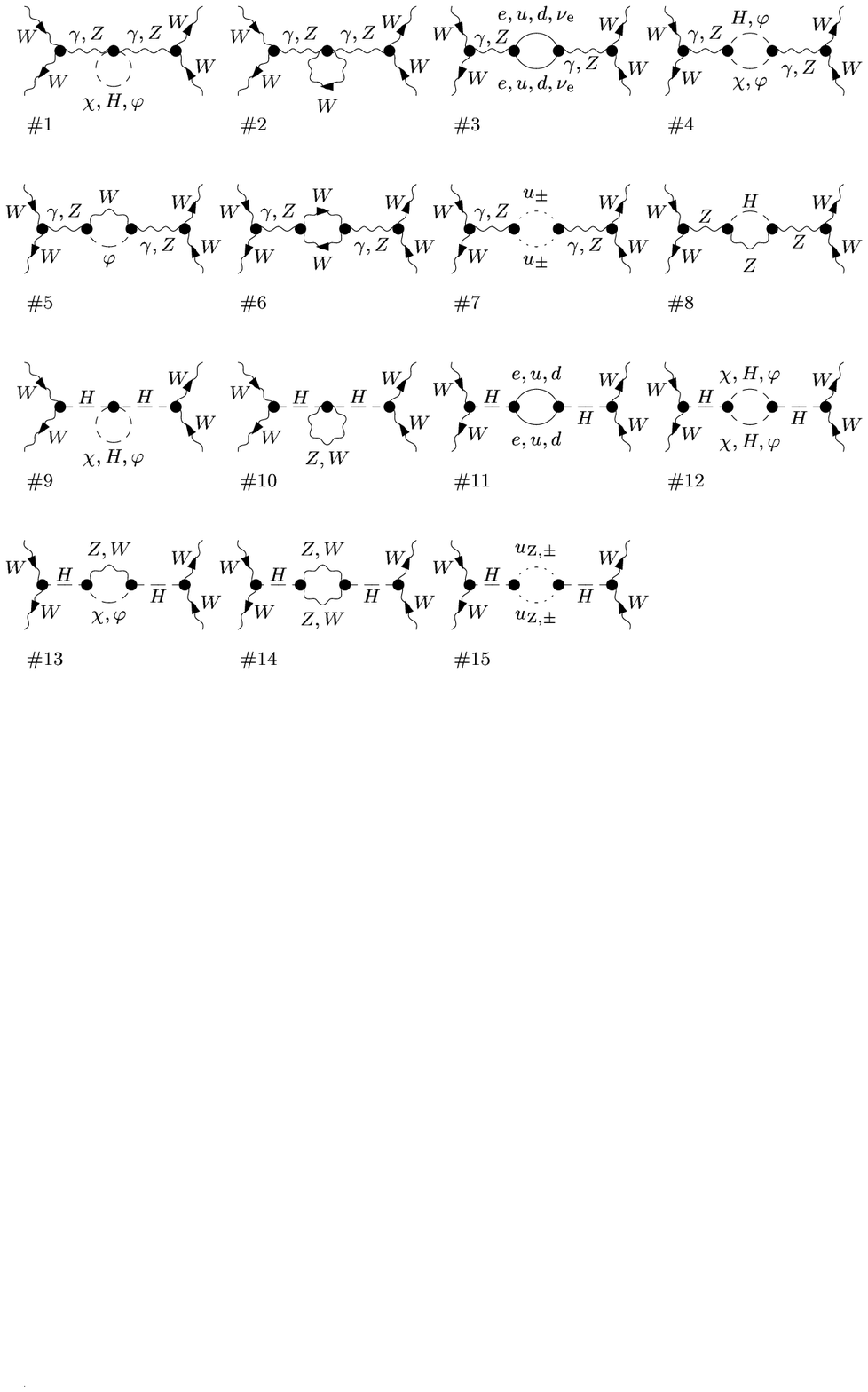}
\end{center}
\vspace*{-2ex}
\caption{\label{fig:selfdiags}The $s$-channel self-energy diagrams.}
\end{figure}

\begin{figure}
\begin{center}
\epsfig{figure=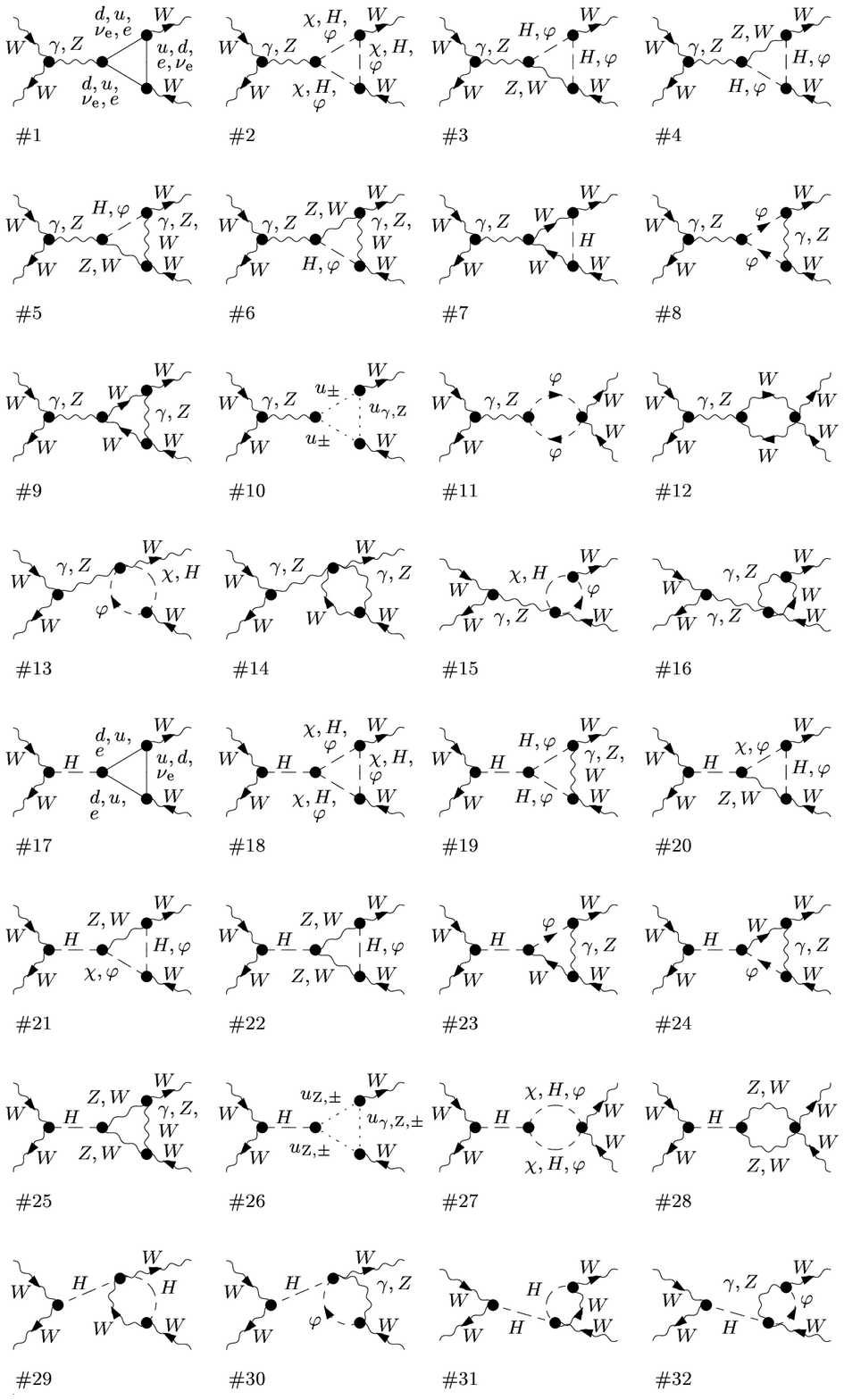}
\end{center}
\vspace*{-2ex}
\caption{\label{fig:vertdiags}The $s$-channel vertex diagrams for the
final-state vertex.}
\end{figure}

In the following we list the Feynman graphs in the conventional
formalism in 't~Hooft--Feynman gauge. To avoid presenting hundreds of
diagrams, similar diagrams have been combined. Each combination of
particle labels on the internal lines corresponds to an extra diagram,
where those have to be omitted that violate charge or lepton-number
conservation or involve one of the following vertices: $\gamma\gamma
HH$, $\gamma\gamma\chi\chi$, $\gamma ZHH$, $\gamma Z\chi\chi$,
$W^+W^-\chi H$, $\chi HH$, $Z\chi\chi$, $ZHH$, $\gamma\chi\chi$,
$\gamma\chi H$, $\gamma HH$, $\gamma\nu\nu$, which do not exist in the
SM. For example, diagram \#1 of \reffi{fig:selfdiags} represents a
$\phi$-loop contribution to the photon self-energy and to the $\gamma
Z$ and $Z\gamma$ mixing, and $\phi$-, $\chi$-, and $H$-loop
contributions to the Z-boson self-energy.  If the charge flow in the
internal lines is not indicated, it can take both directions (\eg in
diagram \#5 in \reffi{fig:selfdiags}) or it is determined from the
charges of the external \PW~bosons (\eg in diagram \#2 in
\reffi{fig:vertdiags}).  In addition, fermionic diagrams are shown
only for the first fermion generation.

The $s$-channel self-energy corrections are shown in
\reffi{fig:selfdiags}. They consist of insertions of the $\PZ$-boson and
photon self-energies and the $\gamma$--$\PZ$-boson mixing energy into
diagram \#1 of \reffi{fig:borndiags} and Higgs-boson self-energy
insertions into diagram \#4 of \reffi{fig:borndiags}. 

The $s$-channel vertex corrections consist of corrections to the 
$\gamma WW$, $ZWW$ and $HWW$ vertices in diagrams \#1 and \#4 of
\reffi{fig:borndiags}. The corrections to the final-state vertices in
these diagrams are displayed in \reffi{fig:vertdiags}.

\begin{figure}
\begin{center}
\begin{tabular}[t]{ll}
\raisebox{12.3cm}{(a)} & 
\epsfig{figure=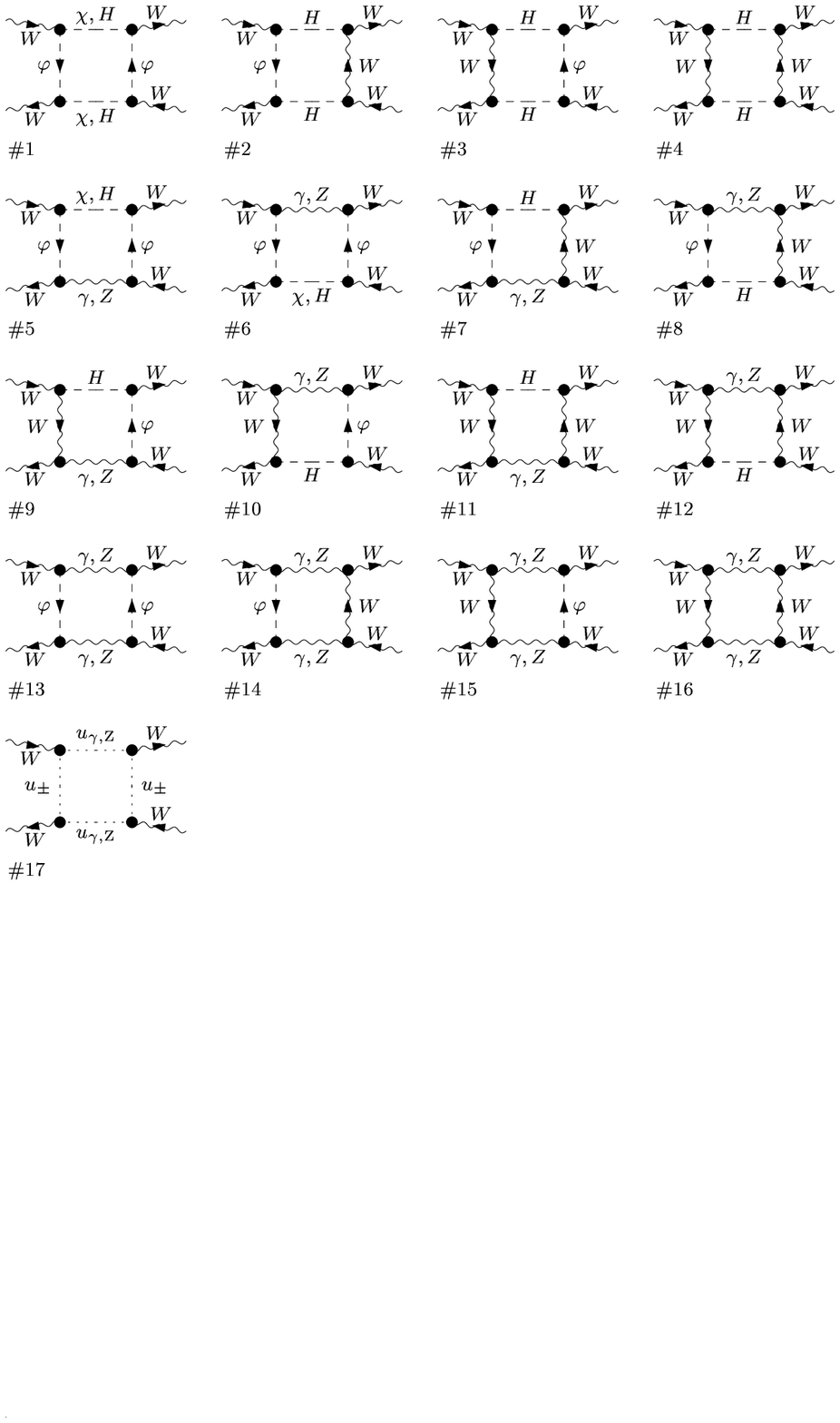} 
\\[1ex]
\raisebox{2.3cm}{(b)} &
\begin{picture}(330,75)(0,-10)
\SetScale{.75}
\SetWidth{1}
\ArrowLine(5,85)(25,65)
\ArrowLine(25,65)(25,25)
\ArrowLine(25,25)(5,5)
\SetWidth{.5}
\DashLine(25,25)(65,25){4}
\DashLine(25,65)(65,65){4}
\SetWidth{1}
\ArrowLine(85,5)(65,25)
\ArrowLine(65,25)(65,65)
\ArrowLine(65,65)(85,85)
\Text(34,-1)[t]{\small (shown above)}
\SetOffset(85,0)
\ArrowLine(5,85)(25,65)
\ArrowLine(25,65)(65,65)
\ArrowLine(65,65)(85,85)
\SetWidth{.5}
\DashLine(25,65)(25,25){4}
\DashLine(65,25)(65,65){4}
\SetWidth{1}
\ArrowLine(85,5)(65,25)
\ArrowLine(65,25)(25,25)
\ArrowLine(25,25)(5,5)
\SetOffset(170,0)
\ArrowLine(5,85)(25,65)
\ArrowLine(25,65)(25,25)
\ArrowLine(25,25)(5,5)
\SetWidth{.5}
\DashLine(25,25)(65,25){4}
\DashLine(25,65)(65,65){4}
\SetWidth{1}
\ArrowLine(85,5)(65,65)
\ArrowLine(65,65)(65,25)
\ArrowLine(65,25)(85,85)
\SetOffset(255,0)
\ArrowLine(5,85)(65,65)
\ArrowLine(65,65)(25,65)
\ArrowLine(25,65)(85,85)
\SetWidth{.5}
\DashLine(25,65)(25,25){4}
\DashLine(65,25)(65,65){4}
\SetWidth{1}
\ArrowLine(85,5)(65,25)
\ArrowLine(65,25)(25,25)
\ArrowLine(25,25)(5,5)
\end{picture}
\end{tabular}
\end{center}
\caption{\label{fig:boxSTUdiags}Genuine box diagrams that appear in 
the $s$, $t$, and $u$ channel. Each diagram in (a) represents actually
four diagrams with different charge flows shown in (b), 
\ie one $s$-channel, one $t$-channel, and two $u$-channel diagrams.}
\end{figure}

\begin{figure}
\begin{center}
\epsfig{figure=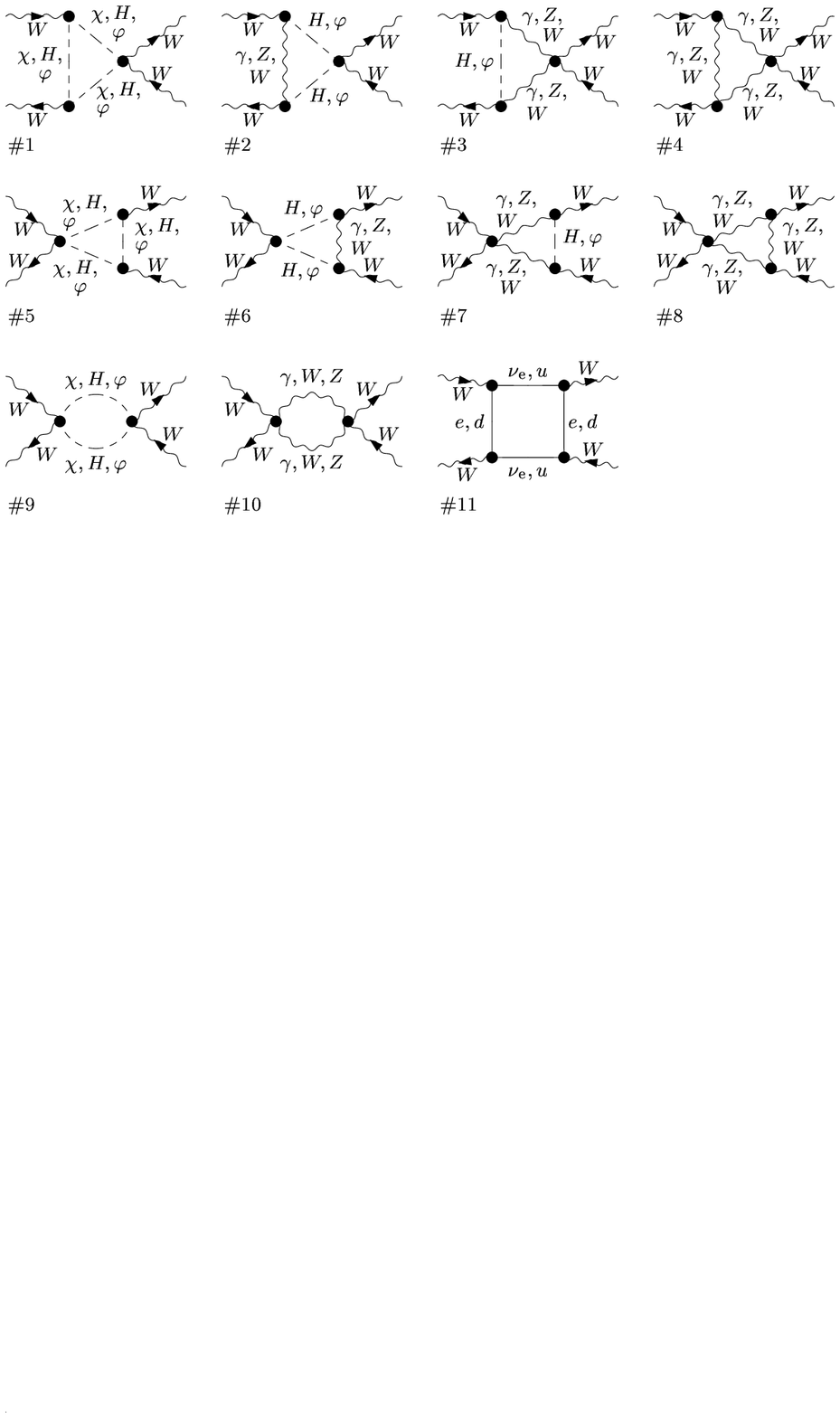} \\[2ex]
\end{center}
\vspace*{-2ex}
\caption{\label{fig:boxSTdiags}Further $s$-channel box diagrams. For every
diagram there exists a $t$-channel analogue (obtained by exchanging the
lower left and upper right leg).}
\end{figure}

\begin{figure}
\begin{center}
\epsfig{figure=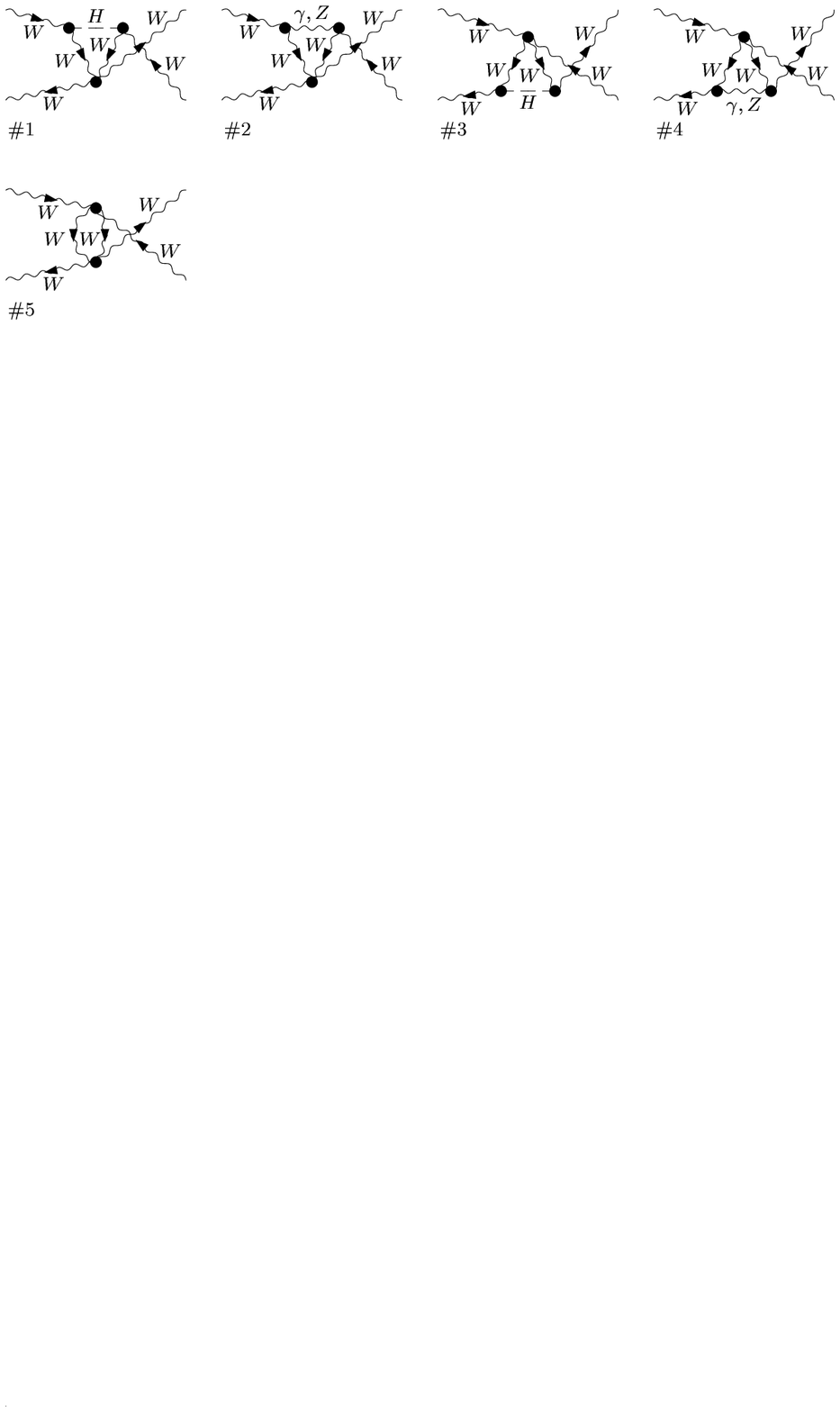}
\end{center}
\vspace*{-2ex}
\caption{\label{fig:boxUdiags}Further $u$-channel box diagrams.}
\end{figure}

The box diagrams are shown in Figs.~\ref{fig:boxSTUdiags}, 
\ref{fig:boxSTdiags}, and \ref{fig:boxUdiags}. While the $t$-channel boxes
are obtained from the $s$-channel boxes via crossing, the $u$-channel
boxes are different owing to the different charge flow. In particular,
there are no fermionic $u$-channel diagrams. For this reason, the Landau
singularity that appears in the corrections to $\ZZZZ$ \cite{DeDH97} does
not show up in $\WWWW$.

Apart from the virtual corrections, one has to take into account real
soft-photon emission from the external \PW~bosons in order to cancel the
infrared (IR) divergences in the charge form factor. The IR divergences
are regularized by a infinitesimal photon mass $\lambda$. In the
soft-photon limit the cross-section for real photon emission is
given by
\begin{equation}
\left(\frac{\d\sigma}{\d\Omega}\right)_{\text{soft}}
=\left(\frac{\d\sigma}{\d\Omega}\right)_{\text{Born}}
\delta_\soft
\end{equation}
with the soft-photon correction factor
\begin{equation}
\label{eq:de_soft}
\delta_\soft=-\frac{e^2}{(2\pi)^3}\int_{k_0\leqslant\Delta E} 
\frac{\d^3k}{2k_0}\sum_{i,j=1}^4
\frac{\pm Q_iQ_j\,(k_i\cdot k_j)}{(k_i\cdot k)(k_j\cdot k)}
\biggr|_{k_0=\sqrt{\vec k^{\,2}+\lambda^2}}\,.
\end{equation}
Here $\Delta E$ is the maximum energy of the emitted photons, $Q_i=\pm 1$
is the charge quantum number of the $i$th W boson, and the sign in front
of the product $Q_iQ_j$ is `$+$' if particles $i$ and $j$ are both either
incoming or outgoing and `$-$' otherwise.

The basic integrals needed for the soft-photon factor have been worked out,
for instance, in \citere{tHV79}. For the case of $\WWWW$ the soft-photon
factor reads
\begin{equation}
\begin{split}
\delta_\soft &= {-}\frac{2\alpha}{\pi}\biggl\{
2\ln\frac{2\Delta E}{\lambda}+\frac Ep\ln\frac{E-p}{E+p} \\
&\qquad -(s-2\MW^2)\;
{\cal I}\!\left(s+\sqrt{s(s-\smash{4\MW^2})}\right) \\
&\qquad -(2\MW^2-t)\;
{\cal I}\!\left(4\MW^2-t+\sqrt{t(t-\smash{4\MW^2})}\right) \\
&\qquad +(2\MW^2-u)\;
{\cal I}\!\left(4\MW^2-u+\sqrt{u(u-\smash{4\MW^2})}\right)
\biggr\},
\end{split}
\end{equation}
where the function ${\cal I}(x)$ is defined as
\begin{equation}
\begin{split}
{\cal I}(x) &= \frac{2(x-2\MW^2)}{x(x-4\MW^2)}\biggl\{
2\ln\frac{2\Delta E}{\lambda}\ln\frac{x-2\MW^2}{2\MW^2} \\
&\qquad +\Li2\!\left(1-\frac{4E(E-p)}{x}\,
\frac{x-2\MW^2}{2\MW^2}\right)
-\Li2\!\left(1-\frac{4E(E-p)}{x}\right) \\
&\qquad +\Li2\!\left(1-\frac{4E(E+p)}{x}\,
\frac{x-2\MW^2}{2\MW^2}\right)
-\Li2\!\left(1-\frac{4E(E+p)}{x}\right)\biggr\}\,.
\end{split}
\end{equation}

\subsection{Leading corrections}

At low energies ($E\sim\MW$) the electroweak radiative corrections
are dominated by the universal corrections associated with the running
of $\alpha$ and the $\rho$ parameter. These effects are typically at
the level of $10\%$. In this energy region no large logarithmic
corrections associated with collinear photons appear in $\WWWW$
because no light external particles are involved.

At high energies ($E\gg\MW$) and for a light Higgs boson, the radiative
corrections are dominated by corrections of the form $(\alpha/\pi) 
\ln(|x|/\MW^2)\ln(|y|/\MW^2)$, $x,y=s,t,u$, which originate from
approximate IR and collinear singularities in the vertex and box
diagrams.%
\footnote{For the process $\Pep\Pem\to\PWp\PWm$ these leading terms
  can be read off from the high-energy expansion given in \citere{eeWWhe}.}
These terms are are of the order of 10 and 20\% at 1 and $10\TeV$,
respectively. At even higher energies these logarithms render the one-loop
results useless and perturbation theory breaks down. In order to obtain
sensible predictions for exclusive cross-sections beyond $10\TeV$ some
kind of resummation of these logarithms is required. When considering more
inclusive quantities and taking into account \PW-boson, \PZ-boson, and
Higgs-boson bremsstrahlung, the $(\alpha/\pi)\ln(|x|/\MW^2)\ln(|y|/\MW^2)$
terms should drop out leaving leading terms of the form $(\alpha/\pi)
\ln(|x|/\MW^2)$.

If the Higgs boson is very heavy, additional large corrections appear.
These corrections are, in particular, important for the scattering of
longitudinal \PW~bosons and have been extensively discussed in the
literature \cite{Pa85,Da89,Ch88}. In the limit $\MH^2\gg s\gg\MW^2$,
terms of the form $(\alpha/\pi\sw^2)(|x|/\MW^2)\ln(\MH^2/|y|)$,
$x,y=s,t,u$, which have been calculated in \citeres{Pa85,Ch88}, dominate
the relative corrections. For $s\gg\MH^2\gg\MW^2$, on the other hand, the
leading relative correction is provided by terms of the form
$(\alpha/\pi\sw^2)(\MH^2/\MW^2)\ln(|x|/\MH^2)$. In this limit also the
two-loop corrections are available \cite{Ri97}. The complete one-loop
corrections proportional to $\MH^2/\MW^2$ to longitudinal \PW-boson
scattering for $s,\MH^2\gg\MW^2$ can be found in \citere{Da89}. Finally,
the leading logarithmic corrections of the form $(\alpha/\pi\sw^2)
\ln(\MH^2/\MW^2)$ in the limit $\MH^2\gg s,\MW^2$ for the scattering of
arbitrarily 
polarized \PW~bosons are given in \citere{Di96}.

\subsection{Checks of the calculation}

Several cross-checks were performed to ascertain the correctness
of our calculation. In particular, we checked
\begin{itemize}
\item the algebraic simplification of the Feynman diagrams 
by comparing the \FO\ results with those obtained from {\sl FeynCalc}
\cite{MeBD91} for selected diagrams (analytically),

\item the numerical evaluation of the Feynman diagrams by comparing 
the evaluation of selected diagrams in {\sl Fortran} and in \mma\
(numerically),

\item the evaluation of the scalar and tensor one-loop integrals by
comparing the results of the {\sl FF} routines with own routines
(numerically),

\item the gauge independence of the results by comparing the results in
conventional formalism and background-field formalism (numerically),

\item the UV finiteness by proving independence of the scale parameter
$\mu$ of dimensional regularization 
(analytically),

\item the IR finiteness by testing independence of the regulator mass
$\lambda$ (numerically).
\end{itemize}
Numerical computations were done using extended precision (quadruple
precision) with a mantissa of approximately 33 digits. The relative
deviation between numerical results from different calculational methods
was less than $10^{-16}$. A double precision calculation is in general
sufficient up to about $10\TeV$. For higher energies the gauge
cancellations become so large that double precision does not give reliable
results any more.


\section{Numerical results for the $\Oa$ corrections}
\label{se:numres}

\subsection{Input parameters and definition of the corrected
  cross-section}

For the calculations the following parameter set is used \cite{PDG}:
\begin{equation}
\begin{aligned}
\alpha^{-1} &= 137.0359895, &
\MZ &= 91.188\GeV,\quad &
\MW &= 80.401\GeV, \\[3pt]
\Me &= 0.51099906\MeV,\quad &
\Mu &= 47.0\MeV,\quad &
\Md &= 47.0\MeV,\quad \\ 
\Mmy &= 105.658389\MeV, &
\Mc &= 1.55\GeV, &
\Ms &= 150\MeV, \\
\Mta &= 1771.1\MeV, &
\Mt &= 175.5\GeV, &
\Mb &= 4.5\GeV. \\
\end{aligned}
\end{equation}
The masses of the light quarks are adjusted such that the experimentally
measured hadronic vacuum polarization is reproduced \cite{Ei95}.
As before, a Higgs-boson mass of $\MH=100\GeV$ is used, and the
integrated cross-sections are evaluated for an angular cut 
$\thcut=10^\circ$.

All cross-sections discussed in the following are calculated by replacing
the matrix element squared in \refeq{eq:diffWQ} by
\begin{equation}
\label{eq:corrcs}
\cabs{\M}^2\to
\cabs{\M_{\text{Born}}}^2(1+\delta_{\text{soft}})+
2\Re\!\left(\M^\ast_{\text{Born}}\delta\M\right)\,,
\end{equation}
where $\delta\M$ is the sum of the one-loop Feynman diagrams,
and $\delta_{\text{soft}}$ the soft-photon correction factor
\refeq{eq:de_soft}. Note that this yields the cross-section including all 
$\Oa$ corrections but neglecting all $\O(\alpha^2)$ corrections. 

In the soft-photon approximation the corrections depend on the soft-photon
cut-off energy $\Delta E$ via logarithms of the form $\ln(\Delta E/E)$. 
For small $\Delta E$, where the soft-photon approximation is valid, these
terms give rise to large corrections, which are, however, cancelled if
hard-photon bremsstrahlung is taken into account. In order to avoid
artificially large soft-photon effects, we have chosen to discard all
terms involving $\ln(\Delta E/E)$ in the corrections.%
\footnote{Of course, these terms have to be reinstalled in a complete
  calculation involving hard-photon  bremsstrahlung.}
The elimination of the cut-off-dependent terms can be simply achieved
by choosing $\Delta E=E$. The corrections defined in this way can be
viewed as a suitable measure of the weak corrections, which cannot be
separated from the electromagnetic corrections on the basis of Feynman
diagrams in a gauge-invariant way \cite{DeDS95}. 

\subsection{Total cross-sections}
\label{sect:totcs1l}

\begin{figure}
\begin{center}
\epsfig{figure=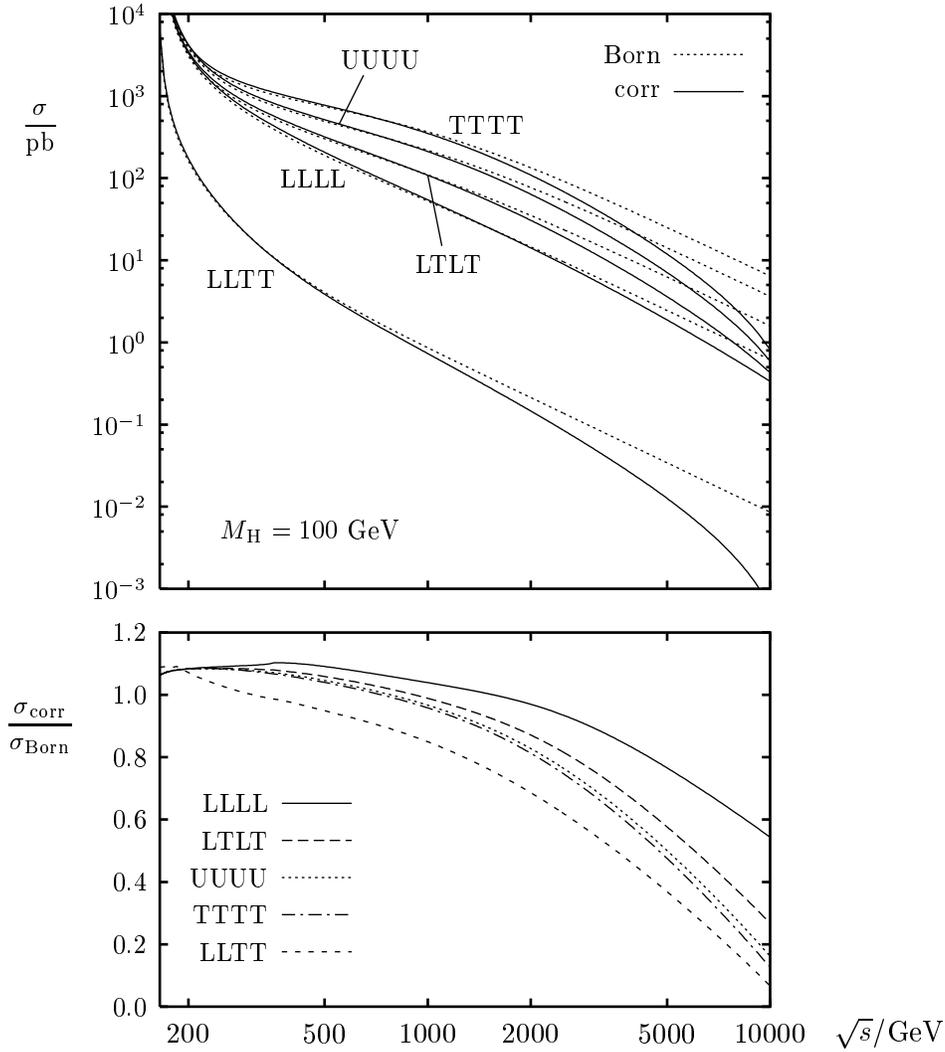}
\end{center}
\caption{\label{fig:onelooptotplots}Upper plot: The integrated
cross-sections for the dominant polarizations in lowest order (Born) and
including the $\Oa$ corrections (corr). Lower plot: Relative corrections
for the same polarizations.}
\end{figure}

In \reffi{fig:onelooptotplots}, the integrated cross-sections in lowest
order and including $\Oa$ corrections as well as the relative $\Oa$
corrections are plotted for the dominant polarizations LLLL, LTLT, LLTT,
TTTT, and UUUU. For small energies the corrections are positive and
roughly 10\%. With increasing energy the corrections become large and
negative and reach the order of the lowest-order cross-sections around
$10\TeV$. The large corrections at high energies originate from
logarithms of the form $\alpha/\pi\ln^2(s/\MW^2)$, which are additionally
enhanced by numerical factors.

\subsection{Differential cross-sections}
\label{sect:1loopdiff}

\begin{figure}
\begin{center}
\begin{tabular}{l}
\epsfig{figure=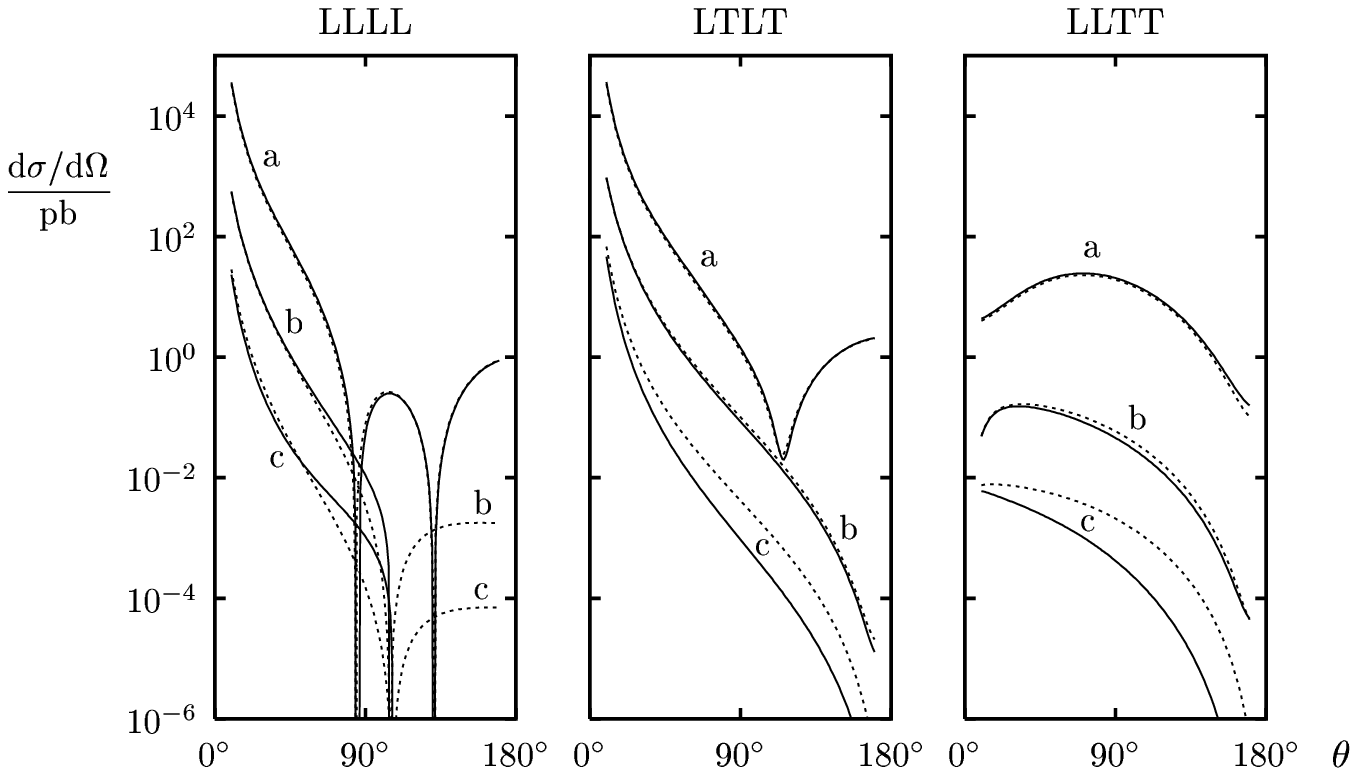} \\[1ex]
\epsfig{figure=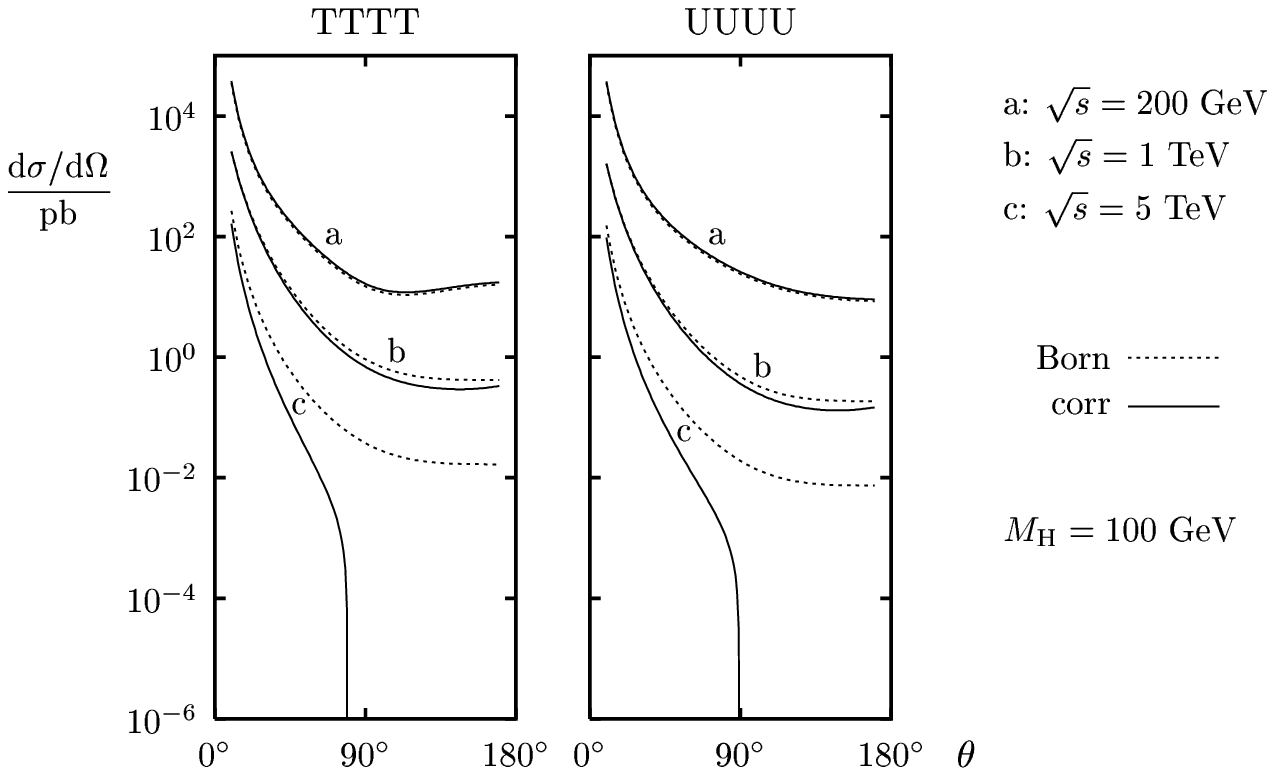}
\end{tabular}
\end{center}
\caption{\label{fig:oneloopdiffplots}The differential cross-sections for 
the dominant polarizations at $\protect\sqrt{s}=200\GeV$, $1\TeV$, and
$5\TeV$ in lowest order (Born) and including the $\Oa$ corrections (corr).
Note that in some regions of phase space the corrected cross-sections
become formally negative and are therefore not visible in these
logarithmic plots.}
\end{figure}

The lowest-order and corrected differential cross-sections for the
dominant polarizations LLLL, LTLT, LLTT, TTTT, and UUUU are shown in
\reffi{fig:oneloopdiffplots} for $\sqrt s=200\GeV$, $1\TeV$, and $5\TeV$.
The corrections are small at low energies for all scattering angles but
get large at high energies, in particular in the backward direction. This
is due to the fact that the contributions of the $u$-channel box diagrams 
to the cross-sections involve terms that behave as $1/|u|$ for
$\MW^2\ll|u|\ll s$ and approach a constant of the order of $1/\MW^2$ for
$|u|\ll\MW^2$, while the lowest-order cross-sections are of order $1/s$
for $|u|\ll s$. Hence the relative corrections in the backward direction
are typically enhanced by a factor $s/|u|$ compared with the corrections
in forward direction or to the total cross-section and can easily reach
100\% at several TeV. Since $\cabs{\delta\M}^2$ is missing in
(\ref{eq:corrcs}), the cross-section can formally become even negative in
this case. By including $\cabs{\delta\M}^2$, the quality of the
predictions can in general be improved, but for $\WWWW$ this leads to
problems, which are discussed in \refse{sect:problems}. Here, we ignore
$\cabs{\delta\M}^2$ and the resulting wrong predictions for high energies
and large scattering angles, since the cross-sections are small in this
region, and, as has been stated above, the total cross-section is
dominated by the forward hemisphere. The large corrections in the backward
direction do not indicate a breakdown of perturbation theory because no
additional enhancement factors $s/u$ should appear in higher orders.

\subsection{Accuracy of the equivalence-theorem predictions}

The calculations of enhanced corrections to longitudinal gauge-boson
scattering \cite{Pa85} usually have been done with the help of the
equivalence theorem (ET) \cite{CoLT74}. When using the ET in higher-order
calculations one has to be careful to include all correction factors that
result from renormalization and amputation \cite{YaY88}. This is
particularly easy within the BFM \cite{DeD96}, where the matrix elements
for external longitudinal vector bosons are directly obtained from the
amputated Green functions with the corresponding would-be Goldstone-boson
fields multiplied with the wave-function renormalization constants of the
gauge bosons. 

We calculated the matrix element for $\pppp$ to one-loop order in the BFM 
and from this the ET prediction for the cross-section for $\WLWLWLWL$
\begin{multline}
\label{eq:WQET}
\left(\frac{\d\sigma}{\d\Omega}\right)_{\text{ET}}=
\frac{1}{64\pi^2s}\biggl[
\bigl|\M_{\text{Born}}^{\pppp}\bigr|^2(1+\de_{\text{soft}}) \\
+2\Re\Bigl\{\bigl(\M_{\text{Born}}^{\pppp}\bigr)^\ast
\delta\M^{\pppp}\Bigr\}\biggr]\,.
\end{multline}
Since $\pppp$ and $\pppp\gamma$ are no physical processes, the
cross-section \refeq{eq:WQET} is not well-defined at finite energies.
The ET is valid only for high energies and guarantees \refeq{eq:WQET}
to provide a sensible cross-section only in the high-energy limit, \ie
up to terms of order $\MW/E$. In fact, at finite energies \refeq{eq:WQET} is
gauge-dependent and the IR singularities between the virtual and
real corrections do not cancel exactly. In the analytic results, which are
obtained for an infinitesimal photon mass $\lambda$, terms proportional to
$(\MW/E)\ln(\lambda^2/s)$ and gauge-dependent terms proportional to
$(\MW/E)$ remain. In order to avoid an enhancement of the IR-singular
terms, we have set the photon mass equal to the energy of the \PW~bosons,
$\lambda=E$, in the numerical evaluation. This ensures that the artificial
IR-singular logarithms do not lead to enhanced contributions. Since also
the soft-photon correction factor corresponding to $\pppp$ is
gauge-dependent we have chosen to use directly the soft-photon correction
factor for $\WWWW$. We stress that despite the gauge-dependences and the
IR-singularities contained in \refeq{eq:WQET}, our results are valid at
the same level as all ET results, \ie up to terms of order $\MW/E$.  

In order to improve the accuracy of the ET calculation we combine the 
lowest-order cross-section for $\WLWLWLWL$ with the $\Oa$ corrections from
$\pppp$ as follows
\begin{multline}
\label{eq:WQmixed}
\left(\frac{\d\sigma}{\d\Omega}\right)_{\text{mixed}}=
\frac{1}{64\pi^2s}\biggl[
\bigl|\M_{\text{Born}}^{\WLWLWLWL}\bigr|^2 
+\bigl|\M_{\text{Born}}^{\pppp}\bigr|^2\de_{\text{soft}}\\
+2\Re\Bigl\{\bigl(\M_{\text{Born}}^{\pppp}\bigr)^\ast
\delta\M^{\pppp}\Bigr\}\biggr]\,.
\end{multline}
In this way the error originating from the use of the ET enters only
via the $\Oa$ corrections. 

\begin{figure}
\begin{center}
\epsfig{figure=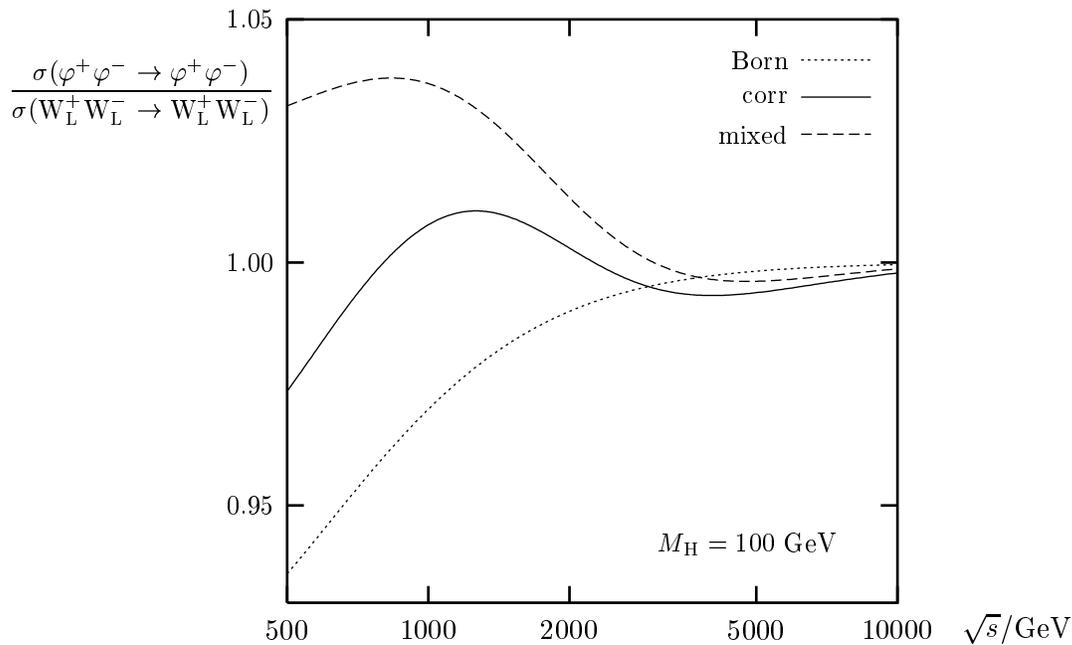}
\end{center}
\caption{\label{fig:ETrel}Ratio of the ET predictions for
$\protect\WLWLWLWL$ and the direct calculation. ``Born'' and ``corr''
label the ratios of the tree-level and $\Oa$-corrected cross-sections,
respectively. For the curve marked ``mixed'' the ET is used only for
the corrections [cf. \refeq{eq:WQmixed}].}
\end{figure}

The accuracy of the ET predictions for the scattering of longitudinal
\PW~bosons is investigated in \reffi{fig:ETrel} and \refta{tab:ETrel}. We
show the lowest-order \cs\ for $\pppp$ normalized to the lowest-order
\cs\ for $\WLWLWLWL$ and the corrected \cs\ calculated from the ET
according to \refeq{eq:WQET} and according to the improved formula
\refeq{eq:WQmixed}, both normalized to the corrected \cs\ for $\WLWLWLWL$. 
For $\sqrt{s}\gtrsim 500\GeV$ the deviations are all below 7\% and become
smaller at higher energies. Above about $2\TeV$ they are less than
1\%. At high energies, where the ET is valid, the approximation
\refeq{eq:WQmixed} is in fact better than the pure ET prediction.
\begin{table}
\renewcommand{\arraycolsep}{2ex}
$$
\begin{array}{|r|r|r|r|}
\cline{2-4}
\multicolumn{1}{c}{} &
\multicolumn{3}{|c|}{
\vrule width 0pt height 4ex depth 3ex
\dfrac{\sigma(\pppp)}{\sigma(\WLWLWLWL)}-1}
\\ \cline{1-1}
\multicolumn{1}{|c|}{\rule{0pt}{2.5ex} \sqrt s} & 
\multicolumn{1}{c|}{\text{Born}} & 
\multicolumn{1}{c|}{\text{corr}} & 
\multicolumn{1}{c|}{\text{mixed}} \\ \hline\hline
500\GeV & -6.4\%     & -2.7\%     &  3.2\%       \\ \hline
1\TeV   & -3.0\%     &  0.8\%     &  3.7\%       \\ \hline
2\TeV   & -1.0\%     &  0.3\%     &  1.3\%       \\ \hline
5\TeV   & -0.2\%     & -0.6\%     & -0.4\%       \\ \hline
10\TeV  & -0.0\%     & -0.2\%     & -0.1\%       \\ \hline
\end{array}
$$
\caption{\label{tab:ETrel}Relative deviations of the ET predictions for
$\protect\WLWLWLWL$.}
\end{table}


\section{Difficulties beyond $\Oa$ accuracy}
\label{sect:problems}

In \refse{se:numres} we found that the $\Oa$ corrections become of
the order of the lowest-order cross-section for high energies. In this
case higher-order corrections are required in order to arrive at a
meaningful prediction. If the importance of the $\Oa$ corrections is due
to a suppression of the lowest-order cross-section, as it happens for
$\WWWW$ at large scattering angles and high energies, it is appropriate to
include the square of the $\Oa$ corrections, \ie to calculate the
cross-section via
\begin{equation}
\label{eq:sigmacorr}
\frac{\d\sigma}{\d\Omega}=\frac 1{64\pi^2 s}
\left|\M_\born+\delta\M\right|^2\,.
\end{equation}
Because the interferences of the genuine $\Oaa$ matrix element with
the lowest-order matrix element are suppressed with the lowest-order
cross-section, this gives a reasonable (leading-order) prediction, 
which is usually sufficient as the cross-section is suppressed. This
recipe has for example been applied in the case of $\ZZZZ$ \cite{DeDH97}.

Since for $\WWWW$ this approach leads to severe problems, it was not
used in this paper.

The first problem is due to the charge of the \PW~boson. It gives rise to
photonic corrections that render the matrix element for the process
$\WWWW$ IR-singular. In order to cancel the IR singularities one has to
add real soft-photon radiation. The cancellation of the IR singularities
takes place order by order. Including $\cabs{\delta\M}^2$ would thus
require to take into account all virtual and real $\O(\alpha^2)$ 
corrections. Since the IR-singular corrections are proportional to the
lowest-order matrix element and thus suppressed, an approximate result
could nevertheless be obtained by including the real $\Oa$ corrections
appropriately when squaring the matrix element. The error introduced by
this effective treatment of the real corrections is of the order of the
neglected $\O(\alpha^2)$ corrections. 

The second, more severe difficulty is closely related to the
instability of the W~boson. When calculating the one-loop diagrams for
$\WLWLWLWL$ one finds that not all terms proportional to $s$ cancel.
The left-over terms proportional to $s$ in the unrenormalized amplitude
can be expressed in terms of the transverse $\PW$ self-energy:
\begin{equation}
\label{eq:SigmaWterms}
\de\M_{\O(s)}^{\text{unren.}}=
\frac{4\pi\alpha}{\sw^2}\cos^2\frac{\theta}{2} \frac{s}{\MW^2}
\frac{\Sigma^W_\rT(\MW^2)}{\MW^2}\,.
\end{equation}
The counter terms, in fact only the \PW-mass counter term $\de\MW^2=
\Re\Sigma^W_\rT(\MW^2)$ (and the derived counter term for the weak mixing
angle $\de\cw^2$), yield a similar contribution with opposite sign. When
adding these contributions, the real part of 
$\de\M_{\O(s)}^{\text{unren.}}$ is exactly cancelled but the imaginary
part survives. As a consequence the imaginary part of the one-loop matrix
element for $\WLWLWLWL$ grows with $s$ and therefore would violate 
unitarity
if taken into account. Because the lowest-order matrix element is real as
long as no Higgs resonance is present, this does not affect the $\Oa$
corrections to the cross-section. However, in higher orders or if the
Higgs-boson width has to be included the imaginary part becomes relevant
and the perturbative evaluation of corrections to $\WLWLWLWL$ for
on-real-mass-shell external \PW~bosons breaks down. This statement holds
for the usual approach where the external \PW~bosons are treated as stable
particles with a real mass, \ie where the corresponding momenta squared are set
equal to the real \PW-boson mass squared.

The reason for the breakdown becomes apparent by noting that the
unitarity-violating term can be written as
\begin{equation}
\label{eq:SigmaImWterms}
\de\M_{\O(s)} =
\frac{4\pi\alpha}{\sw^2}\cos^2\frac{\theta}{2} \frac{s}{\MW^2}
\frac{\i\Im\Sigma^W_\rT(\MW^2)}{\MW^2}
=\i\frac{4\pi\alpha}{\sw^2}\cos^2\frac{\theta}{2}
\frac{s}{\MW^2}\frac{\Gamma_\PW}{\MW}\,,
\end{equation}
i.e.\ it is proportional to the (lowest-order) decay width of the
\PW~boson. This result shows that the problem is due to the instability of
the \PW~bosons, which is inherent in the one-loop calculation. It points
to the fact that $S$-matrix elements for unstable external particles are
not well-defined.

{}From the previous considerations it is clear that the unitarity-violating
terms would cancel if the \PW-boson-mass counter term were complex, i.e.\
if renormalization would also compensate the imaginary part of the
\PW~self-energy. Since the bare \PW-boson mass is real, this can only be
achieved if the renormalized \PW-boson mass (and the renormalized weak
mixing angle) become complex.

Let us consider the matrix element in this renormalization scheme
including $\Oa$ corrections. In the one-loop part and in the counter-term
part, which are of order $\alpha$, the renormalized parameters can be
replaced by bare parameters and all unitarity-violating terms cancel out. 
In the tree-level matrix element, however, the complex renormalized
parameters lead to unitarity-violating terms (if it is calculated for
$k_i^2=\MW^2$). This reflects the fact that a change in renormalization
scheme in an $\Oa$ calculation cannot eliminate $\Oa$ corrections but only
reshuffle them. However, if we evaluate the matrix element for complex
external momenta squared, $k_i^2=\MW^2-\i\MW\Gamma_\PW$, all
unitarity-violating terms cancel in the tree-level part, and the loop and
counter-term parts are not changed (up to terms of order $\alpha^2$)  with
respect to the evaluation with real external momenta squared,
$k_i^2=\MW^2$. As a result, we find that the matrix element for
$\WLWLWLWL$ including $\Oa$ corrections respects unitarity if the squares
of the external momenta are equal to the physical complex masses, but
violates unitarity for real ``on-shell'' external momenta.
This result is independent of the renormalization scheme, and holds, in
particular, in the usual on-shell scheme. Our findings are in accordance
with the fact that gauge invariance and unitarity of $S$-matrix elements
require the momenta of the external particles to be on the mass shell, \ie
on the complex mass shell for unstable particles. They indicate that the
natural generalization of $S$-matrix elements for unstable particles are
the multiple residues at the physical (complex) poles of the external
particles (compare \citere{St64}).

The approach described above requires to consider matrix elements for
complex external momenta squared. However, the momenta of physical
particles are always real, even if these particles appear only as
resonances in some process. Since it is not clear how to relate the
physical real momenta to appropriate complex momenta we have not pursued
this approach any further. The definition of $S$-matrix elements for
unstable external particles remains an unsolved problem.

When calculating the corrections to $\WLWLWLWL$ via the ET, no
unitarity-violating terms appear. The mismatch between the direct
calculation and the ET calculation, \ie the apparent violation of the ET,
can be understood by looking at the derivation of the ET \cite{DeD96}.
The ET results from Ward identities for connected Green functions by
amputation and putting external fields on their mass shell.  Thus, the
validity of the ET demands external \PW~bosons to be on their mass
shell, \ie on the complex mass shell for unstable particles.
For real external on-shell momenta the ET is violated beyond $\Oa$
by terms proportional to the \PW-boson decay width that are enhanced by
factors $E/\MW$ originating from longitudinal polarization vectors.
The failure of the ET is just another symptom of the lack of a
proper description of $S$-matrix elements for unstable particles.

The only fully consistent way to calculate corrections to longitudinal
\PW-boson scattering beyond $\Oa$ seems to be the evaluation of a physical
process, \eg $\Pep\Pem\to\WWWW\to$~4~fermions, or at least the resonant
contributions to this process. This is beyond the scope of this paper.


\section{Conclusions}
\label{se:concl}

The scattering of the weak gauge bosons is an extremely useful tool
for the investigation of the non-abelian gauge sector and the
symmetry-breaking sector of the electroweak interaction. To lowest
order these reactions involve only interactions between gauge and
scalar bosons and, therefore, depend very sensitively on these
couplings. Owing to its large cross-section, a particularly
interesting process in this class is the scattering of charged W
bosons, $\WWWW$.

This process is known in the literature including only the leading
corrections in the limit of high energies and large Higgs-boson masses.
In this paper we have extended these results by calculating the complete
virtual and soft-photonic $\Oa$ corrections to the amplitudes for $\WWWW$
as predicted by the electroweak Standard Model. 

The cross-sections with equal initial- and final-state polarizations
exhibit a $1/t^2$ singularity resulting from the $t$-channel
photon-exchange diagram and are therefore dominant. The corresponding
$\Oa$ corrections are typically of the order of 10\% for energies below
roughly $2\TeV$, grow with energy, and reach the order of the lowest-order
cross-section at several TeV. The corrections become particularly large 
in the backward direction, where the cross-sections are relatively small.
At energies beyond $10\TeV$ the one-loop calculation becomes unreliable
and a resummation of the leading corrections or the inclusion of real 
massive boson emission is required. 

The predictions of the equivalence theorem, which relates the
scattering amplitudes of longitudinally polarized gauge bosons with
those of the unphysical Goldstone-boson fields, have been compared
with the direct calculation at the level of the $\Oa$ corrections and
for a light Higgs boson. For energies above $500\GeV$ the deviations
are below 7\%, above about $2\TeV$ they are less than 1\%.

We have found that the calculation of radiative corrections to the
scattering of on-real-mass-shell longitudinal \PW~bosons, \ie to
$\WLWLWLWL$ in the usual sense, becomes inconsistent beyond $\Oa$. More
precisely, the imaginary part of the one-loop matrix element gets a
unitarity-violating contribution proportional to the \PW-boson width. This
contribution does not affect the $\Oa$ corrections to the cross-section as
long as no Higgs resonance appears. It becomes problematic as soon as the
imaginary part of the matrix element enters an observable, \ie if $\Oaa$
corrections to the cross-section or the finite width of the Higgs boson
are included. This result points to the fact that $S$-matrix elements for
unstable external particles are not well-defined. For a consistent
evaluation, the instability of the \PW~bosons has to be taken into account
properly. This seems to be only possible when considering a full process
with stable initial- and final-state particles that involve $\WLWLWLWL$
as an enhanced contribution. 


\section*{Acknowledgements}

We thank M.~B\"ohm and S.~Dittmaier for useful discussions.


\begin{flushleft}

\end{flushleft}

\end{document}